\documentclass[fullpage]{article}
\usepackage[T1]{fontenc}

\usepackage{booktabs}
\usepackage{hyperref}
\usepackage{authblk}
\usepackage{graphicx}
\usepackage{amsmath}
\usepackage[rgb]{xcolor}
\usepackage[space]{grffile}
\usepackage{subcaption}
\usepackage{tikz}
\usetikzlibrary{patterns,positioning,shapes,arrows,decorations.text}
\usepackage{pgfplots}
\pgfplotsset{compat=newest}
\usepackage{caption}
\usepackage{tabularx}
\newcolumntype{s}{>{\hsize=.125\hsize}X}
\usepackage{enumitem}
\usepackage{multirow}
\usepackage{array}
\usepackage{adjustbox}
\usepackage[ruled]{algorithm2e}
\usepackage{algpseudocode}

\definecolor{candypink}{HTML}{F05672}
\definecolor{inchworm}{HTML}{C7F467}% green
\definecolor{capri}{HTML}{3AC2F8} % blue
\definecolor{yelloworange}{HTML}{FFAA33}
\definecolor{cadetblue}{HTML}{ABB8C4} %gray

\title{\textbf{Finding New Diagnostic Information for Detecting Glaucoma using Neural Networks}}
\author[1,6]{Erfan Noury\thanks{Work done during internship at Matroid.}\thanks{These authors contributed equally.}\textsuperscript{$\mathsection$}}
\author[2]{Suria S. Mannil\textsuperscript{\dag}\textsuperscript{\ddag}}
\author[2]{Robert T. Chang\textsuperscript{\dag}$^\natural$}
\author[3]{An Ran Ran}
\author[3]{Carol Y. Cheung}
\author[4]{Suman S. Thapa}
\author[5]{Harsha L. Rao}
\author[5]{Srilakshmi Dasari}
\author[5]{Mohammed Riyazuddin}
\author[2]{Dolly Chang}
\author[5]{Sriharsha Nagaraj}
\author[3]{Clement C. Tham}
\author[1,7]{Reza Zadeh\textsuperscript{\dag}$^\sharp$}
\affil[1]{Matroid}
\affil[2]{Byers Eye Institute, Stanford University, Palo Alto, CA, United States}
\affil[3]{Department of Ophthalmology and Visual Sciences, The Chinese University of Hong Kong, Hong Kong}
\affil[4]{Tilganga Institute of Ophthalmology, Kathmandu, Nepal}
\affil[5]{Narayana Nethralaya, Bangalore, India}
\affil[6]{University of Maryland at Baltimore County, Baltimore, MD, United States}
\affil[7]{Stanford University, Stanford, CA, United States}
\affil[ ]{\textsuperscript{$\mathsection$}\url{erfan.noury@gmail.com}, \textsuperscript{\ddag}\url{suria1@stanford.edu}, $^\natural$\url{rchang3@stanford.edu}, $^\sharp$\url{reza@matroid.com}}
\date{}                     %% if you don't need date to appear
\begin{document}
\maketitle

\section*{Abstract}

We describe a new approach to automated Glaucoma detection in 3D Spectral Domain Optical Coherence Tomography (OCT) optic nerve scans.
First, we gathered a unique and diverse multi-ethnic dataset of OCT scans consisting of glaucoma and non-glaucomatous cases obtained from four tertiary care eye hospitals located in four different countries.
Using this longitudinal data consisting of multimodal testing including visual field/OCT printouts, fundus photos, and treatment information for better ground truth definition of glaucoma, we achieved state-of-the-art results for automatically detecting Glaucoma from a single raw OCT using a 3D Deep Learning system.
These results are close to human doctors in a variety of settings across heterogeneous datasets and scanning environments.

To verify correctness and interpretability of the automated categorization, we used saliency maps to find areas of focus for the model.
Matching human doctor behavior, the model predictions indeed correlated with the conventional diagnostic parameters in the OCT printouts, such as the retinal nerve fiber  layer.
We further used our model to find new areas in the 3D data that are presently not being identified as a diagnostic parameter to detect glaucoma by human doctors.
Namely, we found that the Lamina Cribrosa (LC) region can be a valuable source of helpful diagnostic information previously unavailable to doctors during routine clinical care because it lacks a quantitative printout.
Our model provides such volumetric quantification of this region.
We found that even when a majority of the RNFL is removed, the LC region can distinguish glaucoma.
This is clinically relevant in high myopes, when the RNFL is already reduced, and thus the LC region may help differentiate glaucoma in this confounding situation.

We further generalize this approach to create a new algorithm called \textsc{DiagFind} that provides a recipe for finding new diagnostic information in medical imagery that may have been previously unusable by doctors.

\section{Introduction}

\begin{figure}[h!]
\centering
\includegraphics[width=\linewidth]{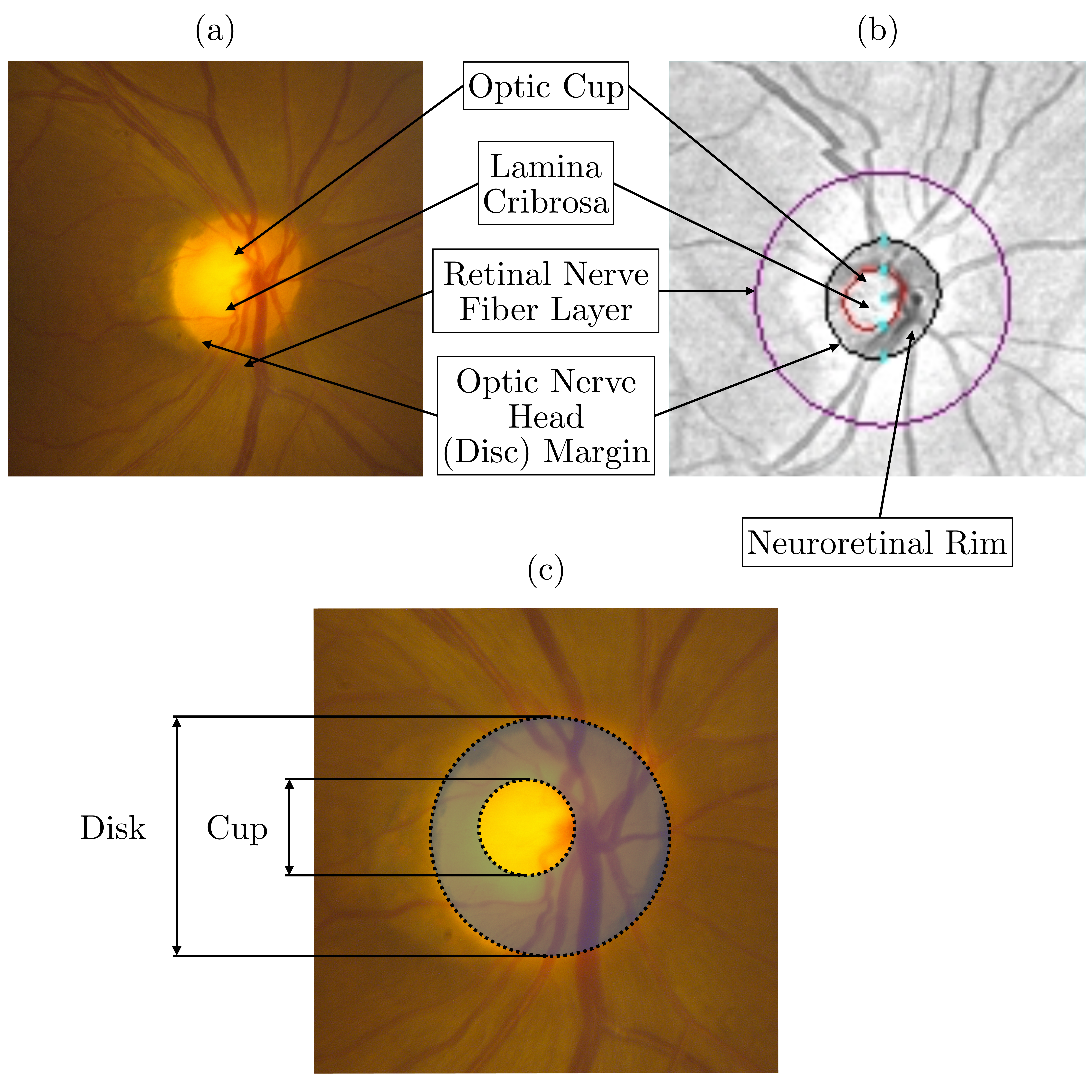}
\caption{Primer on Optic Nerve Head (ONH) Morphology.
\textbf{(a)} Color Fundus Image of the Optic Disc.
\textbf{(b)} Enface OCT image reconstruction of the Optic Nerve Head.
Retinal Nerve Fibers converge at the ONH (known as the optic disc boundary, marked in black in (b)) and then exit the eye as the optic nerve.
The ONH consists of retinal nerve fibers from the Retinal Ganglion Cell axons leading into a central depression known as the optic cup (boundary marked in red in \textbf{(b)}) and a collagenous structure, known as the Lamina Cribrosa, which provides physical support to the exiting axon fibers.
Neuroretinal Rim is the retinal nerve fiber tissue between the border of the cup and the disc.
Optic disc cupping, characterized by progressive neuroretinal rim thinning, is a result of an increased ratio between the optic cup and disc, called the vertical cup-to-disc ratio \textbf{(c)}, a classic feature in glaucoma.
Lamina Cribrosa forms the bottom of the optic cup on the inner surface of the ONH.}
\label{fig:eye}
\end{figure}

Glaucoma, one of the leading causes of irreversible blindness, is a chronic progressive optic neuropathy with characteristic visual field defects matching structural changes, including nerve fiber layer thinning with ganglion cell loss and corresponding optic nerve neuroretinal rim reduction, known commonly as ``cupping''~  \cite{weinreb2004primary,coleman2001glaucomas}.
Currently, the main modifiable risk factor is elevated intraocular pressure (IOP), which, in combination with structural and functioning longitudinal imaging, is one of the main parameters followed during treatment.
Spectral Domain Optical Coherence Tomography (SD-OCT) provides high-resolution cross-sectional imaging of the macula and optic nerve head, in which handcrafted segmentation algorithms applied to a portion of the data produce a quantitative printout for the clinician to interpret based on a normative database.
OCTs operate on the principle of constructive and destructive laser interferometry to estimate a thickness value from the reflected laser light; thus the technology behaves like ultrasound but uses light waves instead of sound waves.
Standardized pathologic glaucomatous structural changes include retinal nerve fiber layer (RNFL) and ganglion cell inner plexiform layer (GCIPL) thinning, but normative cutoffs need to be adjusted by age, ethnicity, refractive error, etc., and thus often  patients are compared to themselves over time to detect glaucomatous progression and confirm disease.
It is also known that glaucoma damage extends deep into the optic nerve head (ONH) at the level of the lamina cribrosa (LC), a network of columns supporting the neuronal axon connections as they traverse from the surface of the retina to the visual cortex of the brain (see \autoref{fig:eye}) \cite{sigal2014recent}. However, only qualitative enhanced depth imaging (EDI) SD-OCT research protocols have been able to visualize these changes in the past and no quantitative printouts exist.

Based on current understanding of high pressure induced glaucoma, biomechanical deformation and remodeling of the ONH leads to posterior displacement of the lamina cribrosa relative to the sclera as well as progressive loss of ganglion cell axons and cell bodies, resulting in RNFL thinning \cite{sigal2014recent}.
Thus it seems reasonable to hypothesize that there is additional information in a standard SD-OCT optic nerve head cube outside of the extracted RNFL which currently is not being tracked clinically, but can be discovered through deep learning as a separate differentiator of glaucoma from normal.

\begin{figure}[h!]
\centering
\begin{tikzpicture}[%
    % common options for blocks:
    block/.style = {
    draw, rectangle, align=center, anchor=north,
    minimum height=0.65cm, inner sep=0.25cm,
    rounded corners=2mm, node distance=2cm, line width=0.25mm},
    arrow/.style ={-latex, line width=0.4mm}]

    \node[block, rounded corners=0mm] (patient) at (0, 0) {Patient};
    \node[block, below of=patient, rounded corners=0mm, fill=white!90!black] (tests) {Patient history, eye exam, and\\multiple diagnostic tests};

    \node[block, below of=tests, xshift=5cm, draw=red!70!black] (abnormal) {Most or all abnormal};
    \node[block, below of=abnormal, fill=red!20, draw=black] (trueglaucoma) {True Glaucoma};
    \node[block, below of=trueglaucoma, draw=blue, node distance=4cm, fill=blue!10] (treatment) {Medical, Surgical,\\or Laser Treatment};

    \node[block, below of=tests, draw=orange!70!black] (halfnormal) {Some abnormal};
    \node[block, below of=halfnormal, fill=orange!15] (suspect) {Suspect};
    \node[block, below of=suspect, draw=blue, fill=blue!10] (monitoring) {Monitoring};
    \node[block, below of=monitoring, xshift=1.5cm, fill=red!5] (refsus) {High-Risk\\Suspect};
    \node[block, below of=monitoring, xshift=-1.5cm, fill=green!5] (nonrefsus) {Low-Risk\\Suspect};

    \node[block, below of=tests, xshift=-5cm, draw=green!70!black] (normal) {None abnormal};
    \node[block, below of=normal, fill=green!20, draw=black] (truenormal) {True Normal};

    \draw[arrow] (patient) -- (tests);

    \draw[arrow] (tests.west) -| (normal.north);
    \draw[arrow] (tests.south) -- (halfnormal.north);
    \draw[arrow] (tests.east) -| (abnormal.north);

    \draw[arrow] (abnormal) -- (trueglaucoma);
    \draw[arrow] (trueglaucoma) -- (treatment);

    \draw[arrow] (halfnormal) -- (suspect);
    \draw[arrow] (suspect) -- (monitoring);
    \draw[arrow, bend left=45] (monitoring.south east) to (refsus.north);
    \draw[arrow, bend left=45] (refsus.north) to (monitoring.south east);
    \draw[arrow, bend left=45] (monitoring.south west) to (nonrefsus.north);
    \draw[arrow, bend left=45] (nonrefsus.north) to (monitoring.south west);

    \draw[arrow] (normal) -- (truenormal);

    \def\myshift#1{\raisebox{1ex}}
    \draw[arrow, postaction={decorate, decoration={text along path, text align=center, text={|\myshift|Progression}}}] (refsus.east) -- ++(0.4cm,0) |- (trueglaucoma.west);

    \def\myshift#1{\raisebox{1ex}}

    \draw[arrow, postaction={decorate, decoration={text along path, text align=center, text={|\myshift|No progression}}}] (nonrefsus.west) -| (truenormal.south);

\end{tikzpicture}
\caption{Glaucoma screening procedure diagram.}
\label{fig:glaucoma}
\end{figure}
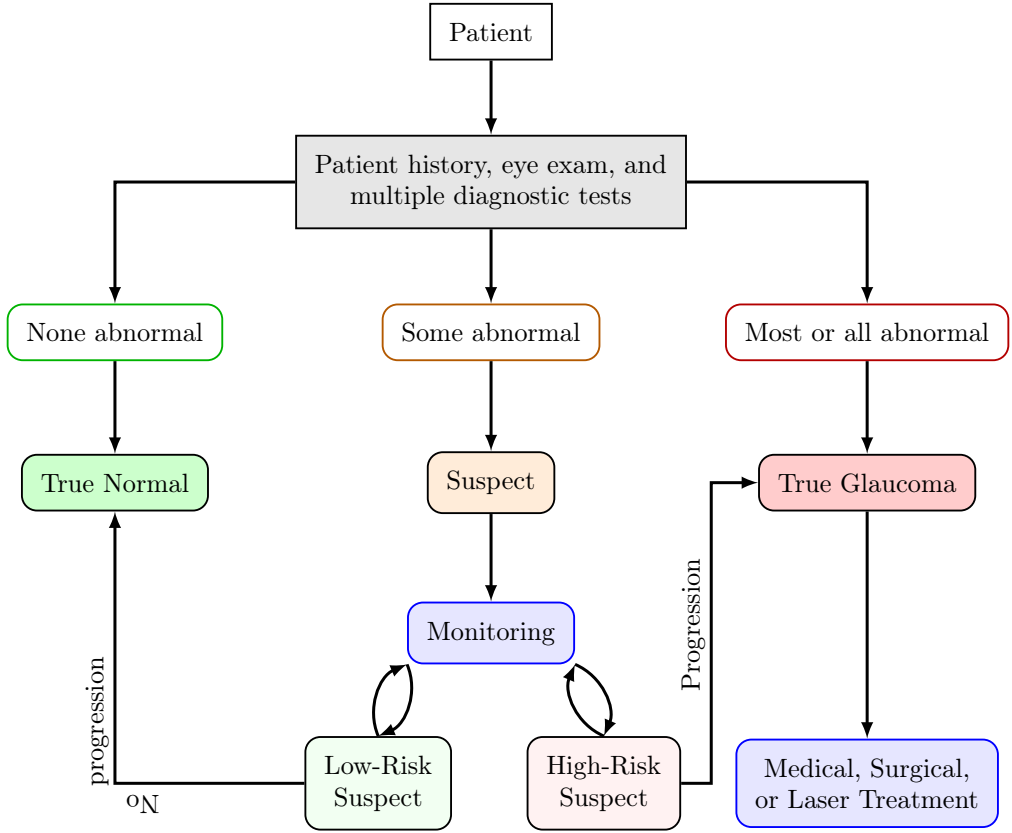

In a typical glaucoma patient evaluation workflow, multiple tests are acquired, similar to the diagram in \autoref{fig:glaucoma}.
This demonstrates the complexity of true glaucoma diagnosis and lack of a single biomarker.
Reaching a consensus ground truth definition of glaucoma on real world data is critical to be able to label an OCT for training an algorithm.
In this work, we concentrate on training a deep learning system which can predict whether an optic nerve head cube scan belongs to the normal (\textit{True Normal}) versus glaucoma (\textit{True Glaucoma}) class based on better ground truth and a hold-out of glaucoma suspects.

To perform this study, data from four different countries was used in the evaluation of the model.
The main data, used for training in our experiments, comes from Byers Eye Institute, Stanford University School of Medicine, while other datasets have been gathered and labeled in Hong Kong, India, and Nepal.
To our knowledge, this is one of the most diverse, multi-ethnic OCT dataset of glaucomatous and non-glaucomatous cases obtained from across four countries.

An important step in effective utilization of artificial intelligence in medicine is the study of its generalization and performance across different datasets from different populations.
This study gives us a glimpse into how the real-world performance of a model for glaucoma can be, when faced with differences in demographics, case inclusion criteria, and subjective international variations in diagnostic criteria for glaucomatous cases.

A deep learning system comprising of a 3D convolutional neural network was used for the experiments.
Due to the large size of the raw cube data, the network was trained on down-sampled 3D cube data to predict one of the \textit{True Normal} or \textit{True Glaucoma} classes.
We observe that the network was able to achieve close to human-level accuracy on two of the four test sets.
To assign a ground truth label to each scan, our ophthalmologists used other clinical data including intraocular pressure, clinical history, OCT printouts, disc photos, and visual field printouts.
However, the only form of information available to the network was the raw 3D OCT scan and a single class label per scan, and we did not have access to other fine-grained ground truth information, e.g. segmentation map, or 2D thickness map.
This was done because we hypothesized there may be additional information within the 3D cube scan that is not being used clinically at the present time.
After training the network, to verify correctness and interpretability of the network, we analyzed the predictions of the model using saliency maps and observed that in most of the \textit{True Glaucoma} cases, the optic nerve head area was highlighted.
Currently, RNFL and GCIPL parameters are used as diagnostic criteria for the presence or absence of glaucoma in an OCT scan interpretation, and the information available from the optic nerve head area which includes the lamina cribrosa region is less well-defined during routine clinical care.
Even though 3D OCT images are readily available in the SD-OCT scanning machine, clinicians often do not have time to view every slice in either the macula or the optic nerve head and thus rely on pre-processed summary printouts.
However, based on the observation that the optic nerve head, including the lamina cribrosa, was often highlighted in the prediction of glaucoma, it seemed this area could help when myopia made the retinal nerve fiber layer thin, and thus we wanted to measure the model's performance on the partial nerve head data alone.
If the measured performance, given only the partial data, was better than a random classifier, it would show that the given area contains relevant diagnostic information and the area of interest can be used in further medical procedures and investigations.
To test the hypothesis in practice, a subset of the scans from the Stanford test set were manually cropped by an ophthalmologist to contain only the optic nerve head area.
However, due to the laborious nature of the manual annotation task, no scans were cropped for training purposes.
Instead, during training, an additional data augmentation step was used to make the resulting model more robust against partial data during evaluation.
We observed that given the optic nerve head, instead of the full scan, as input to the network, the model was able to achieve a better than random classification accuracy, which shows that the optic nerve head area contains diagnostic information that can be used as signals for detecting the presence or absence of Glaucoma, and may have the potential to be used by medical professionals to increase their predictive accuracy.
This new technique of identifying new and informative areas in medical imagery data and devising a new training and evaluation process for assessing the performance of the model given only the data from the area of interest can have ramifications for all of medical imaging.
We further generalize this approach and create an algorithm called \textsc{DiagFind}.
This algorithm can be used for finding new diagnostic information in other medical imagery tasks and can potentially be used for discovering other areas not typically used by doctors, that contain diagnostic information.

In the next section, detailed information about the datasets used in this study are provided.
In Section 4, details of our deep learning system are described and the \textsc{DiagFind} algorithm is introduced.
In Section 5, results from the deep learning system on each test set are provided and analyzed.

\section{Data}

The study adhered to the tenets of the Declaration of Helsinki \cite{world2001world}, and the protocols were approved by the respective institutional review boards of Stanford School of Medicine (United States), Chinese University of Hong Kong (Hong Kong), Narayana Nethralaya Foundation (India), and Tilganga Institute of Ophthalmology (Nepal).
Funding to extract data and store it in a de-identified, encrypted cloud storage was supported by Santen, Inc., Matroid, Inc., and Stanford Global Health Seed Grant by Stanford Center for Innovation in Global Health (CIGH).
Informed consent was waived based on the study's retrospective design, anonymized dataset of OCT images and test data, minimal risk, and confidentiality protections.

\subsection{Data Source}

The current OCT output used in routine glaucoma diagnosis is an RNFL and ONH analysis map (Carl Zeiss Meditec, Inc., Dublin, CA, USA) representing a $6\text{mm}\times 6\text{mm} \times 2\text{mm}$ cube of A-scan data centered over the optic nerve from which a 3.4 mm diameter circle of RNFL data is extracted to create a TSNIT (temporal, superior, nasal, inferior, temporal) 2D map (see Supplementary \autoref{supfig:rnflmap}).
Thickness data from the 2D map is displayed as a four color scale referenced to an age-adjusted normative database.
The normative database for this conventional Cirrus SD-OCT RNFL and ONH map consists of 284 healthy individuals with an age range between 18 and 84 years (mean age of 46.5 years).
Ethnically, 43\% were Caucasian, 24\% were Asian, 18\% were African American, 12\% were Hispanic, 1\% were Indian, and 6\% were of mixed ethnicity.
The refractive error ranged from $- 12.00 D$ to $+8.00 D$~\cite{cirrus}.
Due to the relatively small normative database, there is a significant percentage of false positives from high myopia disc changes or thin RNFL from other non-glaucomatous or artifactual reasons.
One of the difficulties in diagnosing Glaucoma is that there is no single test with a high sensitivity and specificity to confirm the diagnosis which is why OCT alone is not the best label.
Currently, clinicians incorporate the color scale OCT printouts but use a multimodal ground truth label of glaucoma.
This includes clinical examination of the optic nerve head, intra ocular pressure measurement, visual field evaluation, treatment and other relevant clinical history along with OCT RNFL and GCIPL maps to more accurately confirm \textit{True Glaucoma}.

The 3D SD-OCT ONH cube (volume) scans of the training, validation, test, and the external datasets, used in our study, were acquired using Cirrus HD-OCT (Carl Zeiss Meditec, Dublin, CA, USA) according to the optic disc cube scanning protocol.
3D OCT cube (volume) ONH scans of 2202 eyes of 1253 patients evaluated at the Byers Eye Institute, Stanford School of Medicine, from March 2010 to December 2017, were extracted and used for the study.
Prior to labeling as \textit{True Glaucoma} versus \textit{True Normal}, based on chart review, 749 eyes were excluded due to the presence of other ocular pathologies and 93 eyes were excluded due to the presence of OCT artifacts or due to signal strength being less than 3, as per exclusion criteria mentioned below.
61 eyes diagnosed as having preperimetric Glaucoma, 267 eyes diagnosed as suspects (high and low risk) were excluded based on chart review.
20 eyes were excluded after arbitration as described below.
Finally 1012 eyes of 562 patients (2461 scans) were labelled and used for training, validation, and testing.

\subsection{Ground-truth Labeling}

The inclusion criteria were (1) age equal to or older than 18 years old; (2) reliable visual field (VF) tests; and (3) availability of SD-OCT Optic Disc scans.
A reliable visual field report is defined as (a) fixation losses less than 33\%; (b) false positive rate less than 25\%; (c) false negative rate less than 25\%; and (d) no appearance of lid or lens rim artifacts, and no appearance of cloverleaf patterns.
SD-OCT scans with signal strength less than 3 or any artifact obscuring imaging of the ONH, or any artifacts or missing data areas that prevented measuring the thickness of the RNFL at 3.4 mm diameter were excluded from the study.
Artifacts included blink, motion, registration, and mirror artifacts. The reason a signal strength of $\geq 3$ was included is because the entire cube of data was being used and not the results from the machine's segmentation algorithm (which often fails at low signal strength).

\textit{True Glaucoma} was defined as those eyes with glaucomatous disc changes \cite{foster2002definition} on fundus examination, with localized defects on OCT RNFL/GCIPL deviation or sector maps, that correlated with the VF defect which fulfilled the minimum definition of Hodapp-Anderson-Parrish (HAP) glaucomatous VF defect and are on or had intraocular pressure lowering treatment as per chart review \cite{anderson1992automated}. Thus, no pre-perimetric glaucoma were included.
\textit{True Normal} was defined as non-glaucomatous optic disc on fundus exam with no structural defects on OCT RNFL/GCIPL deviation or sector map and normal visual fields, and normal intraocular pressures.
Eyes with optic nerve head pathologies, such as non-glaucomatous optic neuropathy, optic nerve head hypoplasia, or optic nerve pit, and other retinal pathologies such as retinal detachment, age-related macular degeneration, myopic macular degeneration, macular hole, diabetic retinopathy, and arterial and venous obstruction were carefully excluded.

More information about ground-truth labeling are provided in Supplementary \autoref{suptbl:criteria}.

\subsection{Training and Validation}

In total, from the Stanford dataset, 1022 optic nerve scans of 363 eyes from 207 patients with a diagnosis of glaucoma (\textit{True Glaucoma}) (randomly chosen), and 542 scans of 291 eyes from 167 patients of definitive normal (\textit{True Normal}) were included in the training set. % -> Numbers updated
A total of 142 scans of 48 eyes from 27 patients with a \textit{True Glaucoma} annotation, and 61 scans of 39 eyes from 23 patients with a \textit{True Normal} annotation were included in the validation set. % -> Numbers updated

The splitting data into different sets was based on patients, to make sure that scans belonging to each patient are included in only one of the splits and there is no data leakage between different sets.
Each OCT scan over the optic nerve head is a three-dimensional array of size $6\text{mm} \times 6\text{mm} \times 2\text{mm}$ divided into a cube of resolution of $200 \times 200 \times 1024$, with numbers representing the height, width, and depth of the array, respectively.
For the dataset from Stanford, cases were labeled according to the criteria mentioned above by a glaucoma fellowship trained ophthalmologist with more than two years' experience (S.S.M.) based on fundus images, VF, OCT RNFL, GCIPL parameters, and IOP lowering treatment (based on chart review).
In cases where labeling needed arbitration, a senior glaucoma specialist with more than ten years' experience (R.T.C.) reviewed the cases and his diagnoses were considered final.
20 out of 36 conflicting cases were eliminated based on insufficient data on chart review.
To compute inter-grader agreement for diagnosis, a third glaucoma fellowship trained specialist (D.C.) adjudicated the labeling of randomly selected 50 \textit{True Glaucoma} and 50 \textit{True Normal} cases.
Following this, Cohen's $k$ value was calculated.
Inter-grader agreement calculations resulted in a Light's $k$ (arithmetic mean of Cohen's $k$) of 0.8535, considered to represent almost perfect agreement \cite{landis1977measurement}.

\subsection{Test Sets}

Data from the four datasets from four different countries were used in the evaluation of the model.
Test set from Stanford is composed of 694 additional OCT 3D cube images from the Glaucoma Clinic at Stanford University that were annotated after the initial training and validation sets were annotated.
Of those, 241 OCT 3D cube volumes were from 113 eyes (of 66 patients) that were labeled as \textit{True Normal}, and 453 scans of 157 eyes (of 89 patients) were labeled as \textit{True Glaucoma}. % -> Updated numbers
Hong Kong test set consists of 1625 OCT 3D cube images from Chinese University of Hong Kong, with 666 OCT 3D cubes of 196 eyes (of 99 patients) labeled as \textit{True Normal}, and 959 OCT 3D cubes of 277 eyes (of 155 patients), labeled as \textit{True Glaucoma}.
India test set is composed of 672 OCT 3D cube images of ONH from Narayana Nethralaya Foundation, India.
211 scans from 147 eyes of 98 patients were labeled as \textit{True Normal} and 461 OCT 3D cubes from 171 eyes of 101 patients had a \textit{True Glaucoma} annotation.
Finally, Nepal test set contained 380 OCT 3D cube images of ONH from the Tilganga Institute of Ophthalmology, Nepal.
In this dataset, 158 scans from 143 eyes of 89 patients were labeled as \textit{True Normal}, and 222 scans from 174 eyes of 109 patients were labeled as \textit{True Glaucoma}.

For SD-OCT data from the Hong Kong test set, two trained medical students and a postgraduate ophthalmology trainee (with more than 3 years' of experience in Glaucoma) did the initial quality control and then graded the SD-OCT scans into gradable or non-gradable SD-OCT scans, according to the aforementioned criteria.
Two glaucoma specialists then worked separately to label all the eyes with gradable SD-OCT scans into \textit{True Normal}/\textit{True Glaucoma} combined with VF results.
In this dataset glaucoma was defined as RNFL defects on thickness or deviation maps that correlated in position with the VF defect which fulfilled the definition of glaucomatous VF defects \cite{anderson1992automated}.
Most of the images were labelled as \textit{True Normal}/\textit{True Glaucoma} when the two graders arrived at the same categorization separately, but a few disagreeable cases were reviewed by a senior Glaucoma specialist to make the final decision.

For test sets from India and Nepal, glaucoma specialists each with experience of more than 10 years in Glaucoma labeled the cases into \textit{True Glaucoma} and \textit{True Normal}.
Definitions of \textit{True Glaucoma} and \textit{True Normal} in this dataset were similar to those used at Stanford.

\section{Methods}

\subsection{Network Architecture}

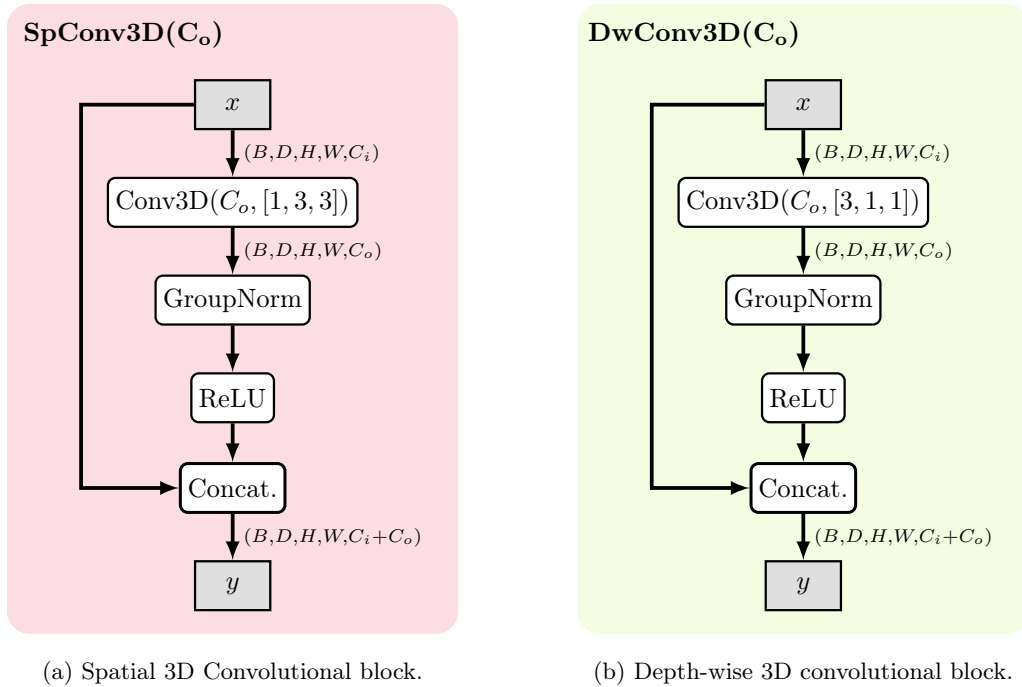
\begin{figure}[h!]
\begin{subfigure}{.5\textwidth}
    \centering
    \pgfdeclarelayer{background}
    \pgfdeclarelayer{foreground}
    \pgfsetlayers{background,main,foreground}

  \begin{tikzpicture}[%
    % common options for blocks:
    block/.style = {
    draw, rectangle, align=center, anchor=north,
    fill=white,
    minimum height=0.65cm, inner sep=0.1cm,
    rounded corners=1mm, node distance=1.3cm, line width=0.3mm},
    arrow/.style ={-latex, line width=0.5mm}]
        \node [block, rounded corners=0mm, fill=white, minimum width=1cm, fill=gray!25!white] (input) at (0, 0) {$x$};
        \node [block, below of=input] (conv) {$\text{Conv3D}(C_o,[1,3,3])$};
        \node [block, below of=conv] (norm) {GroupNorm};
        \node [block, below of=norm] (relu) {ReLU};
        \node [block, below of=relu, node distance=1.2cm, line width=0.4mm] (concat) {Concat.};
        \node [block, below of=concat, rounded corners=0mm, fill=white, minimum width=1cm,
        fill=gray!25!white] (output) {$y$};

        \path (input.north)+(-1.45cm,0.6cm) node (label)[align=center] {\textbf{SpConv3D($\mathbf{C_o}$)}};

        \begin{pgfonlayer}{background}
            \path (input.north)+(-3cm,1cm) node (tl) {};
            \path (output.south)+(+3cm,-0.3cm) node (br) {};

            \path[fill=candypink!20,rounded corners=3mm] (tl) rectangle (br);
        \end{pgfonlayer}

        \def\myshift#1{\raisebox{1ex}}
        \draw [arrow] (input) to node [right] {$\scriptstyle (B, D, H, W, C_{i})$} (conv);
        \draw [arrow] (conv) to node[right] {$\scriptstyle (B, D, H, W, C_{o})$} (norm);
        \draw [arrow] (norm) to (relu);
        \draw [arrow] (relu) to (concat);
        \draw [arrow] (concat) to node[right] {$\scriptstyle (B, D, H, W, C_i + C_o)$} (output);
        \draw [arrow] (input.west) -- ++(-1.5cm, 0) |- (concat.west);
    \end{tikzpicture}
  \caption{Spatial 3D Convolutional block.}
  \label{fig:spatialconv3d}
\end{subfigure}%
\begin{subfigure}{.5\textwidth}
  \centering
    \pgfdeclarelayer{background}
    \pgfdeclarelayer{foreground}
    \pgfsetlayers{background,main,foreground}
  \begin{tikzpicture}[%
    % common options for blocks:
    block/.style = {
    draw, rectangle, align=center,
    anchor=north, fill=white,
    minimum height=0.65cm, inner sep=0.1cm,
    rounded corners=1mm, node distance=1.3cm, line width=0.3mm},
    arrow/.style ={-latex, line width=0.5mm}]
        \node [block, rounded corners=0mm, fill=white, minimum width=1cm, fill=gray!25!white] (input) at (0, 0) {$x$};
        \node [block, below of=input] (conv) {$\text{Conv3D}(C_o, [3, 1, 1])$};
        \node [block, below of=conv] (norm) {GroupNorm};
        \node [block, below of=norm] (relu) {ReLU};
        \node [block, below of=relu, node distance=1.2cm, line width=0.4mm] (concat) {Concat.};
        \node [block, below of=concat, rounded corners=0mm, fill=white, minimum width=1cm, fill=gray!25!white] (output) {$y$};

        \path (input.north)+(-1.45cm,0.6cm) node (label)[align=center] {\textbf{DwConv3D($\mathbf{C_o}$)}};

        \begin{pgfonlayer}{background}
            \path (input.north)+(-3cm,1cm) node (tl) {};
            \path (output.south)+(+3cm,-0.3cm) node (br) {};

            \path[fill=inchworm!20, rounded corners=3mm] (tl) rectangle (br);
        \end{pgfonlayer}

        \def\myshift#1{\raisebox{1ex}}
        \draw [arrow] (input) to node [right] {$\scriptstyle (B, D, H, W, C_{i})$} (conv);
        \draw [arrow] (conv) to node[right] {$\scriptstyle (B, D, H, W, C_{o})$} (norm);
        \draw [arrow] (norm) to (relu);
        \draw [arrow] (relu) to (concat);
        \draw [arrow] (concat) to node[right] {$\scriptstyle (B, D, H, W, C_i + C_o)$} (output);
        \draw [arrow] (input.west) -- ++(-1.5cm, 0) |- (concat.west);
    \end{tikzpicture}
  \caption{Depth-wise 3D convolutional block.}
  \label{fig:depthconv3d}
\end{subfigure}
\caption{Building blocks of the dense convolutional blocks used in the convolutional neural network.}
\label{fig:convlayers}
\end{figure}

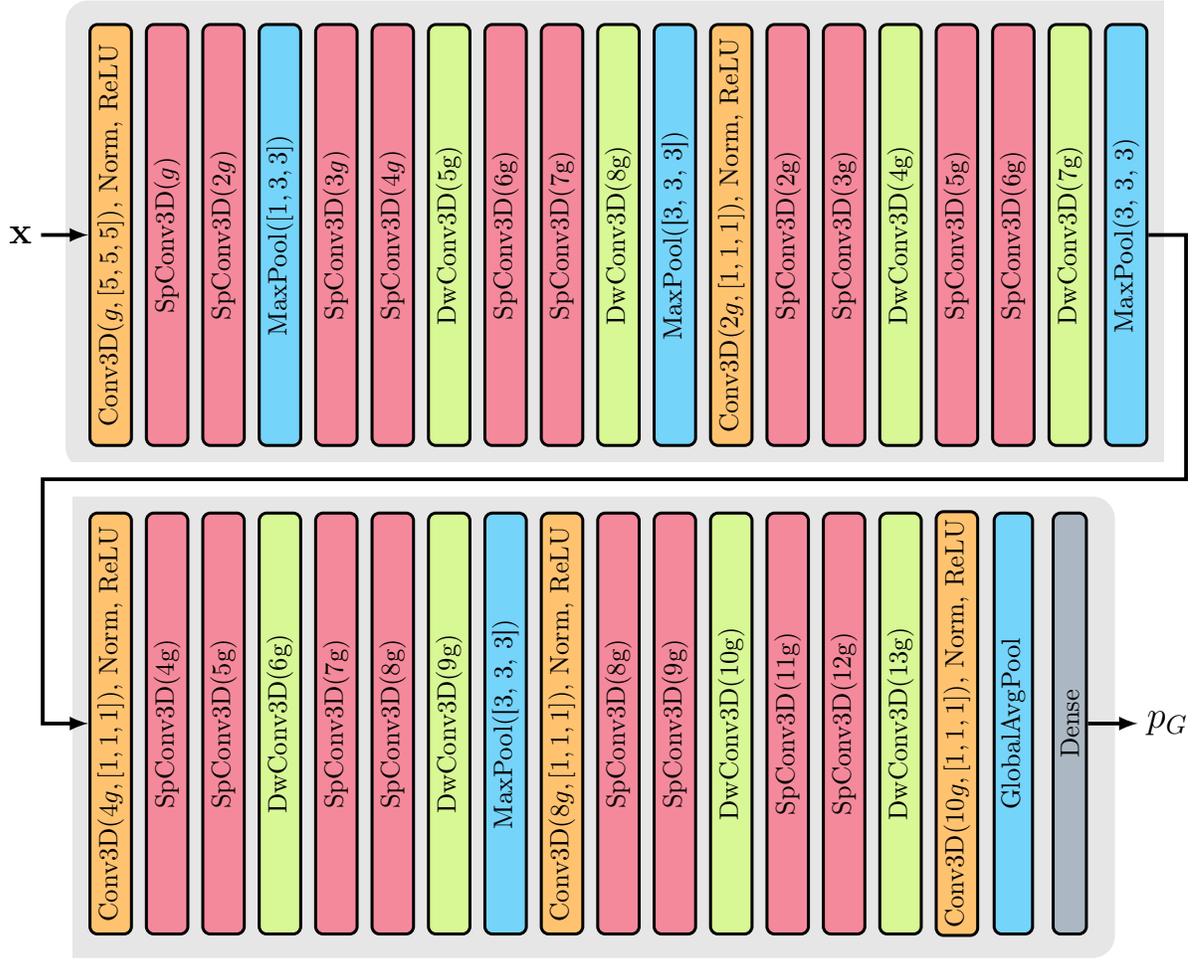
\begin{figure}[h!]
\centering
    \pgfdeclarelayer{background}
    \pgfdeclarelayer{foreground}
    \pgfsetlayers{background,main,foreground}
\begin{tikzpicture}[%
% common options for blocks:
block/.style = {
draw, rectangle, align=center, anchor=north,
fill=white,
minimum height=5.6cm, inner sep=0.1cm,
rounded corners=1mm, node distance=0.75cm, line width=0.4mm},
arrow/.style ={-latex, line width=0.5mm}]
    \node [] (input) at (0, 0) {\Large $\mathbf{x}$};

    \node [block, right of=input, fill=yelloworange!70, node distance=1.2cm] (conv1) {\rotatebox{90}{Conv3D$(g, [5, 5, 5])$, Norm, ReLU}};

    \node [block, right of=conv1, fill=candypink!70] (spconv2) {\rotatebox{90}{SpConv3D$(g)$}};

    \node [block, right of=spconv2, fill=candypink!70] (spconv3) {\rotatebox{90}{SpConv3D$(2g)$}};

    \node [block, right of=spconv3, fill=capri!70] (mp4) {\rotatebox{90}{MaxPool$([1,3,3])$}};

    \node [block, right of=mp4, fill=candypink!70] (spconv5) {\rotatebox{90}{SpConv3D($3g$)}};

    \node [block, right of=spconv5, fill=candypink!70] (spconv6) {\rotatebox{90}{SpConv3D($4g$)}};

    \node [block, right of=spconv6, fill=inchworm!70] (dwconv7) {\rotatebox{90}{DwConv3D(5g)}};

    \node [block, right of=dwconv7, fill=candypink!70] (spconv8) {\rotatebox{90}{SpConv3D(6g)}};

    \node [block, right of=spconv8, fill=candypink!70] (spconv9) {\rotatebox{90}{SpConv3D(7g)}};

    \node [block, right of=spconv9, fill=inchworm!70] (dwconv10) {\rotatebox{90}{DwConv3D(8g)}};

    \node [block, right of=dwconv10, fill=capri!70] (mp11) {\rotatebox{90}{MaxPool([3, 3, 3])}};

    \node [block, right of=mp11, fill=yelloworange!70] (conv12) {\rotatebox{90}{Conv3D$(2g, [1, 1, 1])$, Norm, ReLU}};

    \node [block, right of=conv12, fill=candypink!70] (spconv13) {\rotatebox{90}{SpConv3D(2g)}};

    \node [block, right of=spconv13, fill=candypink!70] (spconv14) {\rotatebox{90}{SpConv3D(3g)}};

    \node [block, right of=spconv14, fill=inchworm!70] (dwconv15) {\rotatebox{90}{DwConv3D(4g)}};

    \node [block, right of=dwconv15, fill=candypink!70] (spconv16) {\rotatebox{90}{SpConv3D(5g)}};

    \node [block, right of=spconv16, fill=candypink!70] (spconv17) {\rotatebox{90}{SpConv3D(6g)}};

    \node [block, right of=spconv17, fill=inchworm!70] (dwconv18) {\rotatebox{90}{DwConv3D(7g)}};

    \node [block, right of=dwconv18, fill=capri!70] (mp19) {\rotatebox{90}{MaxPool(3, 3, 3)}};

    % second row
    \node [block, below of=conv1, fill=yelloworange!70, node distance=6.5cm] (conv20) {\rotatebox{90}{Conv3D$(4g, [1, 1, 1])$, Norm, ReLU}};

    \node [block, right of=conv20, fill=candypink!70] (spconv21) {\rotatebox{90}{SpConv3D(4g)}};

    \node [block, right of=spconv21, fill=candypink!70] (spconv22) {\rotatebox{90}{SpConv3D(5g)}};

    \node [block, right of=spconv22, fill=inchworm!70] (dwconv23) {\rotatebox{90}{DwConv3D(6g)}};

    \node [block, right of=dwconv23, fill=candypink!70] (spconv24) {\rotatebox{90}{SpConv3D(7g)}};

    \node [block, right of=spconv24, fill=candypink!70] (spconv25) {\rotatebox{90}{SpConv3D(8g)}};

    \node [block, right of=spconv25, fill=inchworm!70] (dwconv26) {\rotatebox{90}{DwConv3D(9g)}};

    \node [block, right of=dwconv26, fill=capri!70] (mp27) {\rotatebox{90}{MaxPool([3, 3, 3])}};

    \node [block, right of=mp27, fill=yelloworange!70] (conv28) {\rotatebox{90}{Conv3D$(8g, [1, 1, 1])$, Norm, ReLU}};

    \node [block, right of=conv28, fill=candypink!70] (spconv29) {\rotatebox{90}{SpConv3D(8g)}};

    \node [block, right of=spconv29, fill=candypink!70] (spconv30) {\rotatebox{90}{SpConv3D(9g)}};

    \node [block, right of=spconv30, fill=inchworm!70] (dwconv31) {\rotatebox{90}{DwConv3D(10g)}};

    \node [block, right of=dwconv31, fill=candypink!70] (spconv32) {\rotatebox{90}{SpConv3D(11g)}};

    \node [block, right of=spconv32, fill=candypink!70] (spconv33) {\rotatebox{90}{SpConv3D(12g)}};

    \node [block, right of=spconv33, fill=inchworm!70] (dwconv34) {\rotatebox{90}{DwConv3D(13g)}};

     \node [block, right of=dwconv34, fill=yelloworange!70] (conv35) {\rotatebox{90}{Conv3D$(10g, [1, 1, 1])$, Norm, ReLU}};

     \node [block, right of=conv35, fill=capri!70] (gap) {\rotatebox{90}{GlobalAvgPool}};

     \node [block, right of=gap, fill=cadetblue] (fc) {\rotatebox{90}{Dense}};

     \node [right of=fc, node distance=1.3cm] (output) {\Large $p_G$};

    \draw [arrow] (input) to (conv1);
    \draw [arrow] (fc) to (output);
    \draw [arrow] (mp19.east) -- ++(0.5cm, 0) -- ++(0, -3.25cm) -- ++(-15.2cm, 0) |- (conv20.west);

    \begin{pgfonlayer}{background}
        \path (conv1.north)+(-0.6cm,0.3cm) node (tl) {};
        \path (mp19.south)+(+1cm,-0.3cm) node (br) {};
        \path (mp19.south)+(0.5cm,-0.2cm) node (cbr) {};
        \begin{scope}
        \clip (conv1.north)+(-1cm,1cm) rectangle (cbr);
        \path[fill=gray!20, rounded corners=3mm] (tl) rectangle (br);
        \end{scope}

        \path (conv20.north)+(-1cm,0.2cm) node (bottom_tl) {};
        \path (fc.south)+(0.6cm,-0.3cm) node (bottom_br) {};
        \path (conv20.north)+(-0.5cm,0.2cm) node (clip_bottom_tl) {};
        \path (fc.south)+(1cm,-1cm) node (clip_bottom_br) {};
        \begin{scope}
            \clip (clip_bottom_tl) rectangle (clip_bottom_br);
            \path [fill=gray!20, rounded corners=3mm] (bottom_tl) rectangle (bottom_br);
        \end{scope}
    \end{pgfonlayer}
\end{tikzpicture}
\caption{Architecture of the neural network used in the present study. \textbf{Norm} indicates Group Normalization, and $g$ is a hyperparameter which was set to 16. \textbf{SpConv3D} and \textbf{DwConv3D} are defined in \autoref{fig:spatialconv3d} and \autoref{fig:depthconv3d}, respectively.}
\label{fig:densenetarch}
\end{figure}

A 3D convolutional neural network similar to the classification network of De Fauw \textit{et al.}~\cite{de2018clinically} is used in our experiments (\autoref{fig:densenetarch}).
This network uses multiple layers of dense convolutional blocks \cite{iandola2014densenet}.
Each dense convolutional block consists of one 3D spatial convolutional block (\autoref{fig:spatialconv3d}) followed by a 3D depth-wise convolutional block (\autoref{fig:depthconv3d}).
Each convolutional block applies a convolutional operation, followed by group normalization \cite{wu2018group} and ReLU non-linearity to the input, and the output is concatenated to the input of the convolutional block along the channel axis.
Number of channels in a convolutional layer is defined as a multiple of $g$, which is called growth rate in DenseNet \cite{iandola2014densenet} architecture.
All convolutional layers have a stride of 1, and max pooling stride was set to 2 for dimensions that had a larger than 1 window size.

\begin{figure}[h!]
\begin{subfigure}{.5\textwidth}
  \centering
  \includegraphics[width=\linewidth]{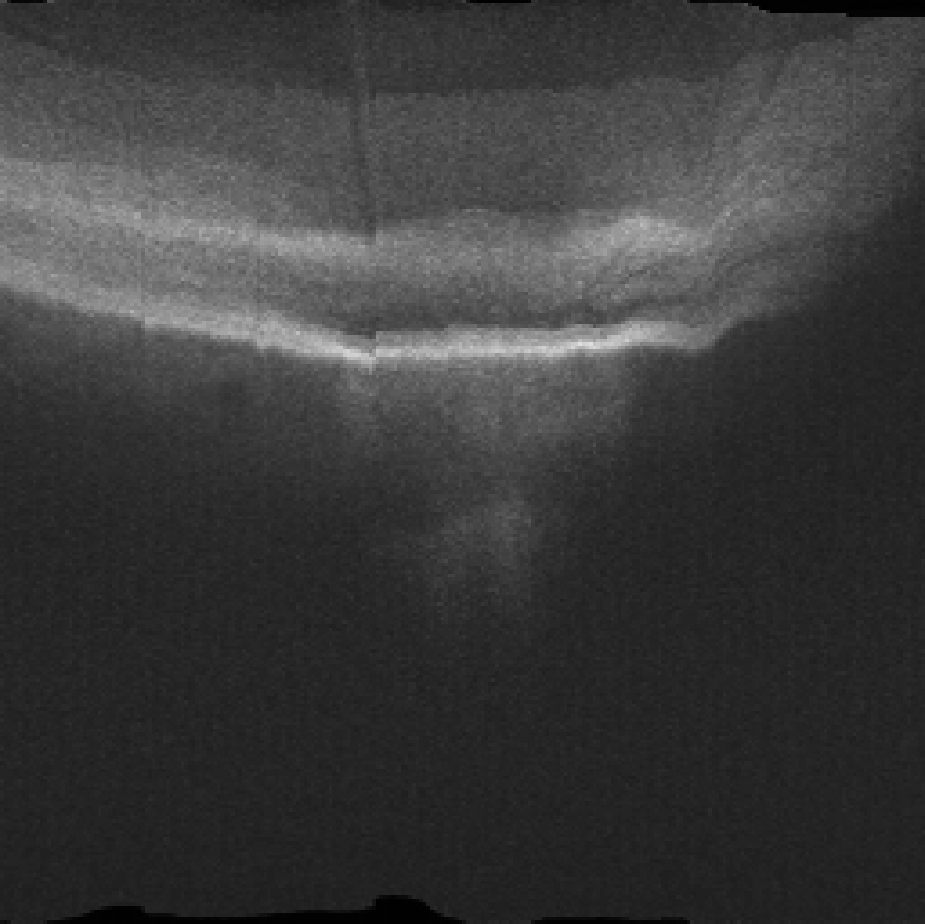}
\end{subfigure}%
\begin{subfigure}{.5\textwidth}
  \centering
  \includegraphics[width=\textwidth]{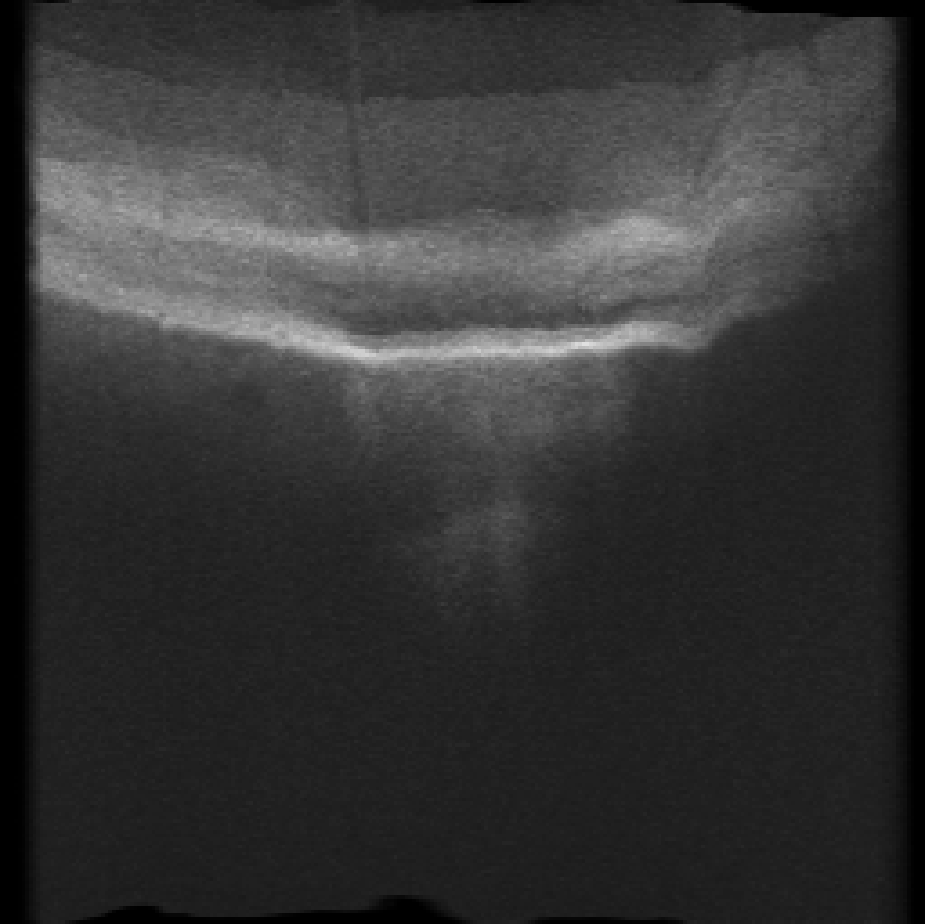}
\end{subfigure}
\\
\\
\begin{subfigure}{.5\textwidth}
  \centering
  \includegraphics[width=\textwidth]{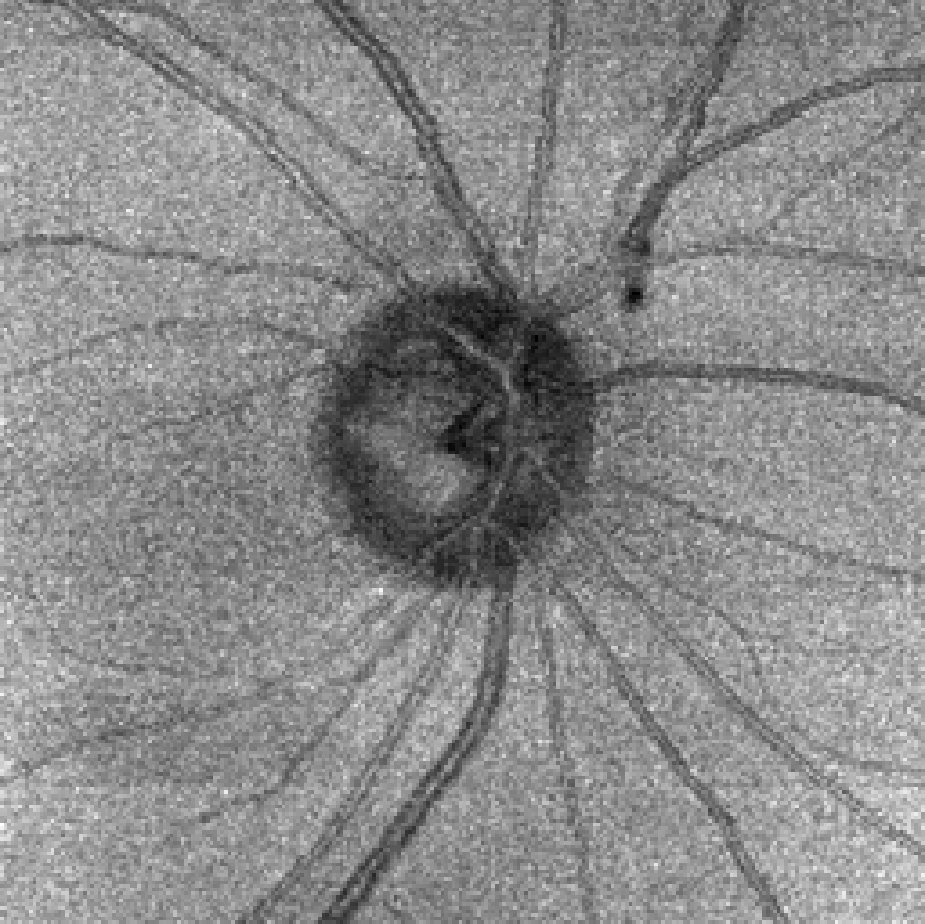}
  \caption{}
\end{subfigure}%
\begin{subfigure}{.5\textwidth}
  \centering
  \includegraphics[width=\textwidth]{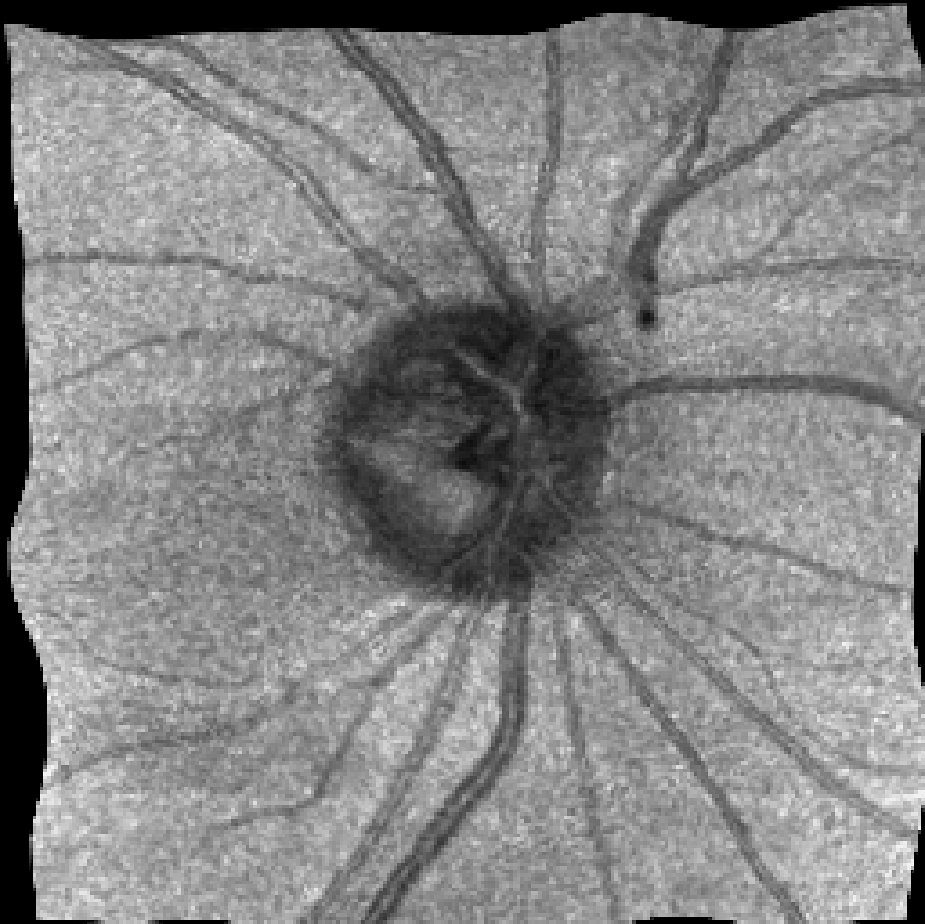}
  \caption{}
\end{subfigure}%
\caption{\textbf{(a)} Original OCT scans. \textbf{(b)} Elastic Deformation applied to the OCT scans.
Darker regions are tissues in the eye that are less transparent against the light beamed to the eye.}
\label{fig:augmentation}
\end{figure}

To increase the amount of effective training data, random flipping and dense elastic deformations were used as data augmentation during training (see \autoref{fig:augmentation}).
Adam optimizer with weight decay \cite{loshchilov2017fixing} was used for training.
After training, model checkpoint with the best results on the validation set was selected as the final model.

\subsection{Finding Areas of Interest}

After training, to get better insight into the predictions of the model, we used saliency methods to try to interpret how the model made its predictions.
For this purpose, the Grad-CAM saliency method \cite{selvaraju2017grad} was used.
We observed that for most of the \textit{True Glaucoma} predictions, lamina cribrosa from the optic nerve head was highlighted by the saliency method, while for majority of the \textit{True Normal} predictions, the retinal layer was highlighted.
In Glaucoma diagnosis, the OCT retinal nerve fiber layer thickness is measured and used to predict whether a patient has glaucoma or not, while other areas of the OCT scan are usually ignored and not considered in the diagnosis process.
However, the highlighting of the lamina cribrosa area can be a new biomarker that potentially can be used by medical professionals as an additional signal in glaucoma detection.
Our goal is for our model to be able to provide a quantification of this diagnostic signal.

\begin{figure}[h!]
\centering
\includegraphics[width=0.6\linewidth]{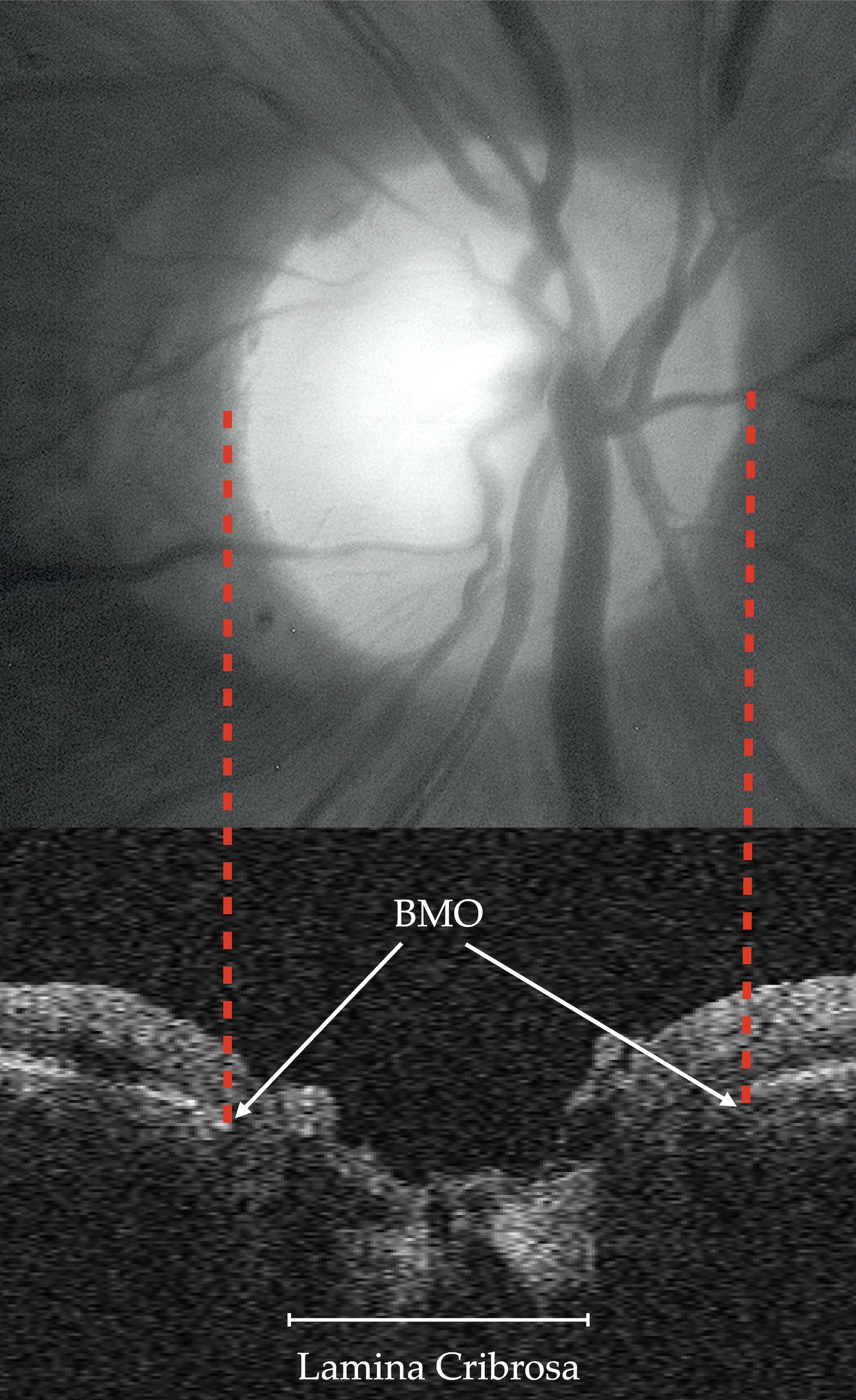}
\caption{Histologically, the optic nerve head is a three-layered opening through which the axons of the retinal ganglion cells (RGCs) pass; the innermost layer is Bruch's membrane opening (BMO), the middle layer is the choroidal opening, and the third layer is the scleral canal opening.
BMO is a distinctly identifiable anatomical structure on spectral-domain optical coherence tomography (SD-OCT); thus, BMO is considered an anatomically accurate and reliable landmark in evaluating the disc margin in glaucoma \cite{chauhan2013clinical}.}
\label{fig:bmo}
\end{figure}

\begin{figure}[h!]
\begin{subfigure}{\textwidth}
    \centering
\includegraphics[width=\linewidth]{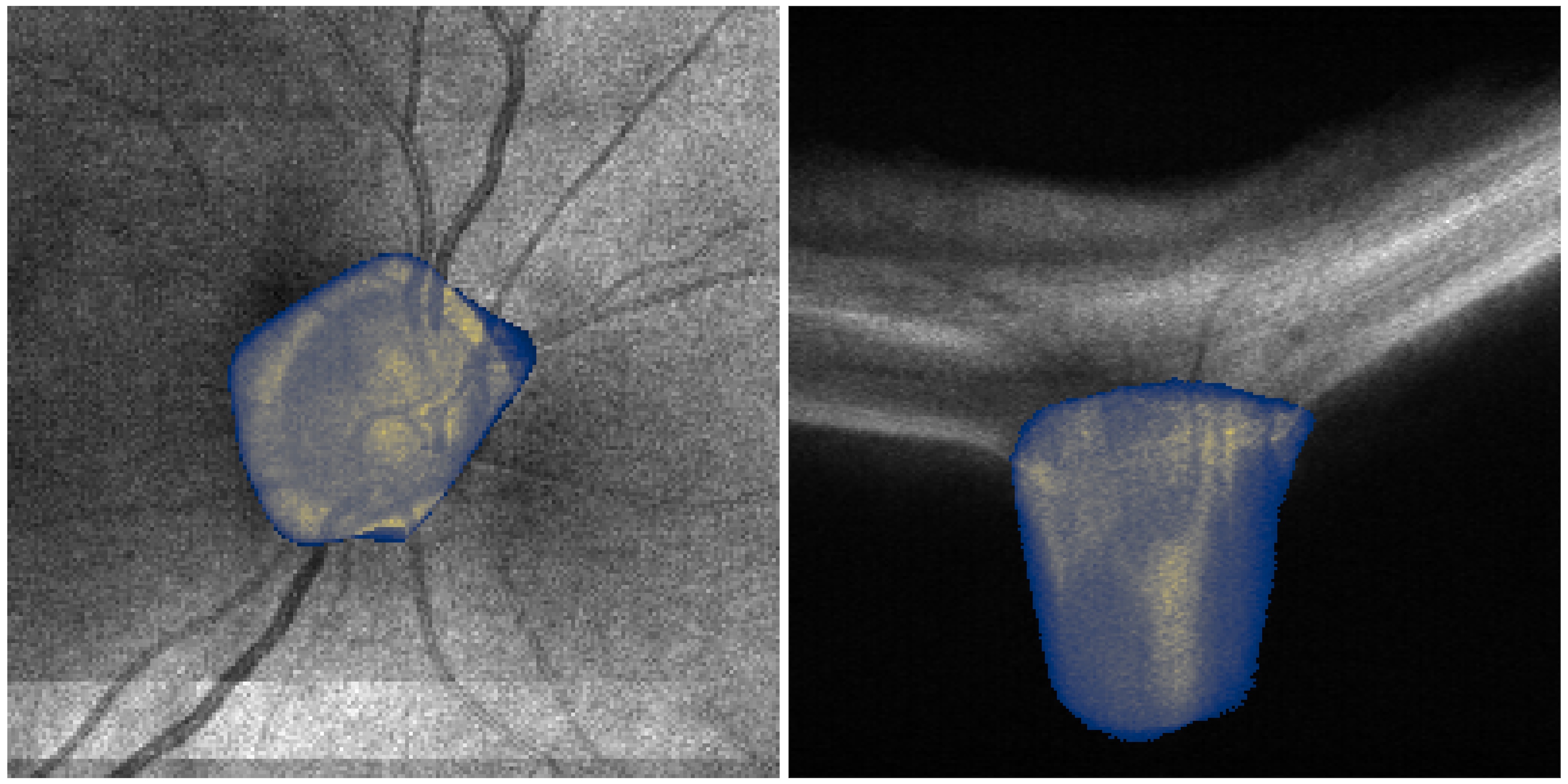}
\caption{}
\end{subfigure}%
\\
\begin{subfigure}{\textwidth}
    \centering
    \includegraphics[width=\linewidth]{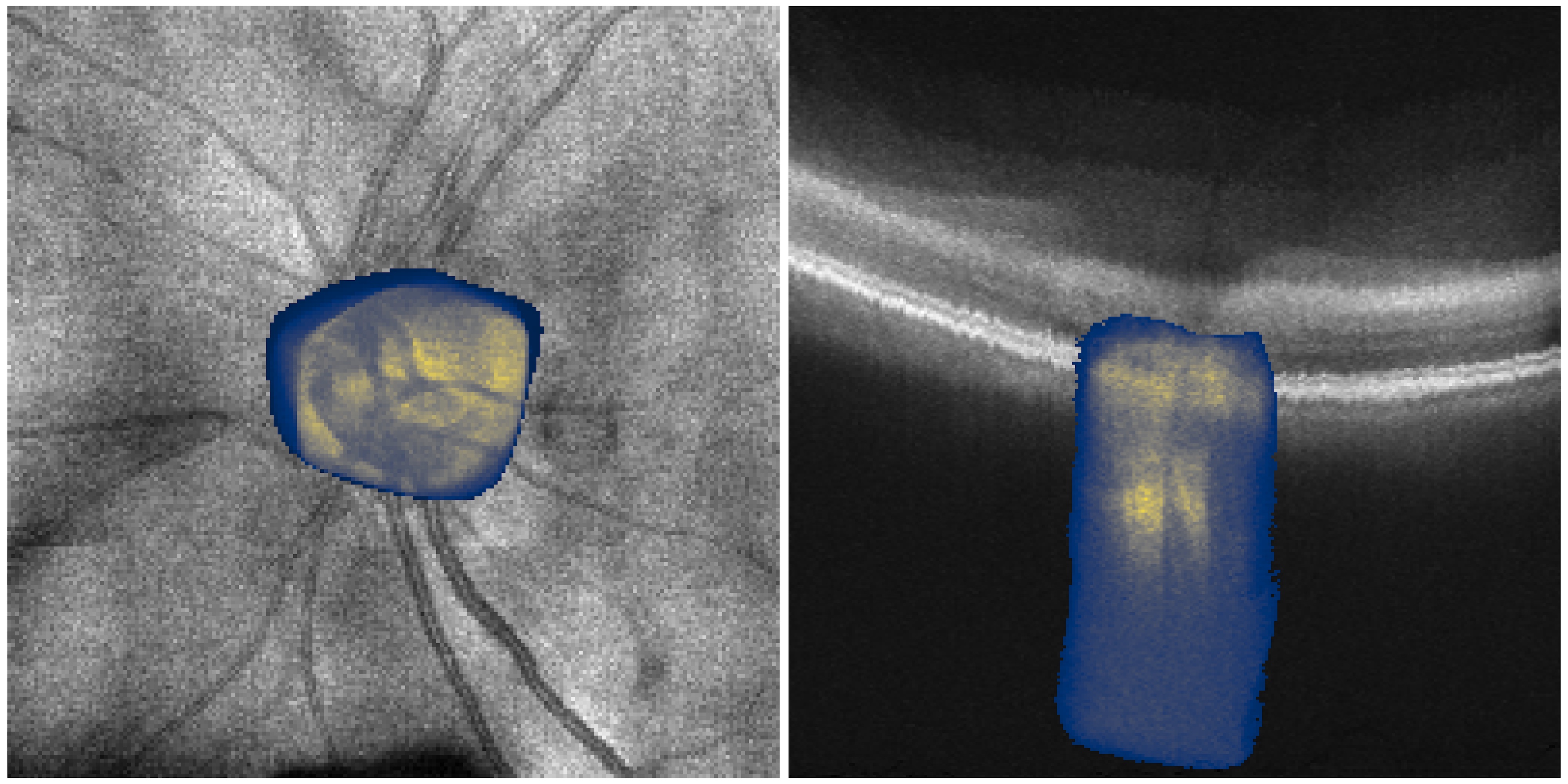}
    \caption{}
\end{subfigure}
\caption{Visualization of the cropped scans, overlaid on the un-cropped scans. Top row \textbf{(a)} shows a \textit{True Normal} scan and the bottom row \textbf{(b)} shows a \textit{True Glaucoma} scan.}
\label{fig:croppedscans}
\end{figure}

To test whether the optic nerve head area of the scan contains any diagnostic information, we devised a new experiment.
Cropping of the OCT images on a small subset of scans was done by glaucoma fellowship trained ophthalmologist (SSM) to only include the optic nerve head, basically creating a 3D mask for the optic nerve head area.
For cropping the scans we used a software known as 3D Slicer \cite{fedorov20123d} which is an open-source software platform for biomedical image informatics, image processing, and three-dimensional visualization.
We cropped optic nerve head region by identifying the appropriate zone of the image in the expected location relative to BMO (see \autoref{fig:bmo}) for a border of dark/light junction at the typical position of the anterior and posterior LC position identified by consolidating and connecting, and identified  individual positions of likely target regions under additional three-dimensional visualisation of axial scans.
Examples of cropped scans are shown in \autoref{fig:croppedscans}.

Our hypothesis was that if a glaucoma detection network can achieve a performance better than a random classifier, given only the cropped scan, then it would show that the optic nerve head area indeed includes informative signals for glaucoma detection.
Remember the cohort includes real world scans even with lower signal strength.

Since manually cropping 3D OCT volumes is laborious and time-consuming, we only annotated 100 OCT scans with equal number of \textit{True Normal} and \textit{True Glaucoma} cases.
Therefore, we were neither able to train our network on cropped data nor searching over hyperparameters was possible.
To solve the first issue, we applied extra cropping data augmentation to make the model more robust against partial data.
In this data augmentation, we randomly selected a smaller volume, and set the values outside the volume to zero.
To mitigate the latter issue, we trained the best performing model from random initialization with the additional data augmentation and used it to get numbers on the cropped scan test set.

We think that the method we used for finding new areas in the OCT scans that contain useful diagnostic information is a general method and can be applied to other problems in medical imaging.
Therefore, we generalize and describe this method, which we call \textsc{DiagFind} in the next section.

\subsection{\textsc{DiagFind}}

The \textsc{DiagFind} (Algorithm \ref{alg:diagfind}) method for finding new areas with diagnostic information in medical imagery tasks consists of multiple steps that are described in \autoref{alg:diagfind}.

\begin{algorithm}
\caption{\textsc{DiagFind}}\label{alg:diagfind}
\begin{algorithmic}[1]
\State Train a neural network on a medical imagery classification task.
\State Utilize saliency methods to find areas of potential sensitivity, and confirm these areas are useful by consulting a domain expert (e.g. a glaucoma-specialized ophthalmologist for this paper)
\State Further refine these areas of sensitivity to those that correlate with a a diagnostic label for which the model is being trained.
\State Redo training, while utilizing a cropping data augmentation that crops the focus onto the areas of sensitivity.

\State Manually crop a number of evaluation data points to the area of interest and evaluate and measure the performance of the model on the cropped data.

\State If the resulting performance of the model is non-trivial, it shows that the identified area contains useful diagnostic information for the given medical imagery problem, since model has no input other than the area of interest.
\end{algorithmic}
\end{algorithm}

If this an area of sensitivity can be positively identified using \textsc{DiagFind}, it can be further analyzed to uncover any causal relations (stronger than the initial perceived correlation) between the model prediction and the newly identified area of interest.
This way the resulting model can also be used to quantify the importance of the area of interest in identifying the diseases in medical imagery problems.

\section{Results}

Area Under the (Receiver Operating) Curve (AUC), sensitivity, specificity, and F1 scores have been used to quantify the performance of the models on the test sets.
In a binary classification model, different discrimination threshold values will result in different values of precision and recall, due to changing values of true positive, true negative, false positive, and false negative.
The Area Under the (Receiver Operating) Curve summarizes the performance of the binary classifier for different values of discrimination threshold.
AUC is also a measure of the probability of the binary classifier giving a random positive sample a higher probability of belonging to the positive class compared to a random negative data point \cite{fawcett2006introduction}.

Sensitivity and specificity are statistical measures that are used to quantify the performance of binary classification models.
In clinical settings, sensitivity measures the percentage of people that have a disease that have also been detected to have the disease by the binary classifier.
Similarly, specificity is a measure of the percentage of people without the disease that have been detected to not have the disease by the binary classifier.
To compute sensitivity and specificity, we used a discrimination threshold from the validation set, such that the resulting predictions would have maximum F2 score, giving more weight to recall than to precision, to have smaller number of false negative predictions.

Demographic background of the training, validation, and test sets are presented in Supplementary Tables \ref{suptbl:stanfordtrainingdemography}, \ref{suptbl:stanfordvaldemography}, and \ref{suptbl:stanfordtestdemography}, respectively.
The demographic data includes age, gender, and ethnicity distribution, visual field mean deviation (MD), and mean refractive error as these are parameters known to affect the OCT cube tissue thicknesses independent of glaucoma.
Note that for some patients, demographic data was incomplete and therefore, aggregate numbers do not necessarily add up to the dataset size.
Demographic information for the Hong Kong, India, and Nepal are presented in Supplementary Tables \ref{suptbl:hkdemography}, \ref{suptbl:indiademography}, and \ref{suptbl:nepaldemography}, respectively.

Among the \textit{True Glaucoma} cases in the training, validation, and test sets from Stanford, there was no significant difference in the average age ($p > 0.005$), but the average age of patients in Hong Kong, India, and Nepal datasets were significantly lower than the training dataset from Stanford ($p < 0.005$).
There was significant difference in the mean refractive error between the \textit{True Glaucoma} and \textit{True Normal} subsets in the Stanford data compared to data from Hong Kong, India, and Nepal ($p < 0.005$).
The dataset from Stanford had significantly higher degrees of myopia compared to the rest of the datasets.
The distribution of cases according to severity of refractive error is shown in Supplementary \autoref{suptbl:myopiaseverity}.
There is significantly higher percentage of severe myopia cases in the \textit{True Glaucoma} subset in the data from Stanford and Hong Kong compared to data from India and Nepal.
Also there is significantly higher number of severe myopia in the \textit{True Normal} subset of the Stanford data compared to Hong Kong, India, and Nepal datasets ($p < 0.005$).
There was no significant difference in severity of glaucoma between the training and validation sets ($p = 0.0724$), and Stanford test set ($p = 0.2709$), Hong Kong test set ($p = 0.035$) and Nepal test set ($p = 0.0369$), while it was significant compared to the India test set ($p < 0.005$).
The percentage of severe glaucoma cases in the India data was significantly higher ($p < 0.005$) compared to data from other sources (Supplementary \autoref{suptbl:severity}).
Severity distribution of datasets from United States, Hong Kong, India, and Nepal are shown in Supplementary \autoref{suptbl:severity}.
Details of additional clinical information such as cup-to-disc ratio, IOP, gender distribution, pattern standard deviation (PSD), and visual field index (VFI) are shown in Supplementary \autoref{suptbl:additional}.

\begin{figure}[h!]
\centering
\includegraphics[width=\linewidth]{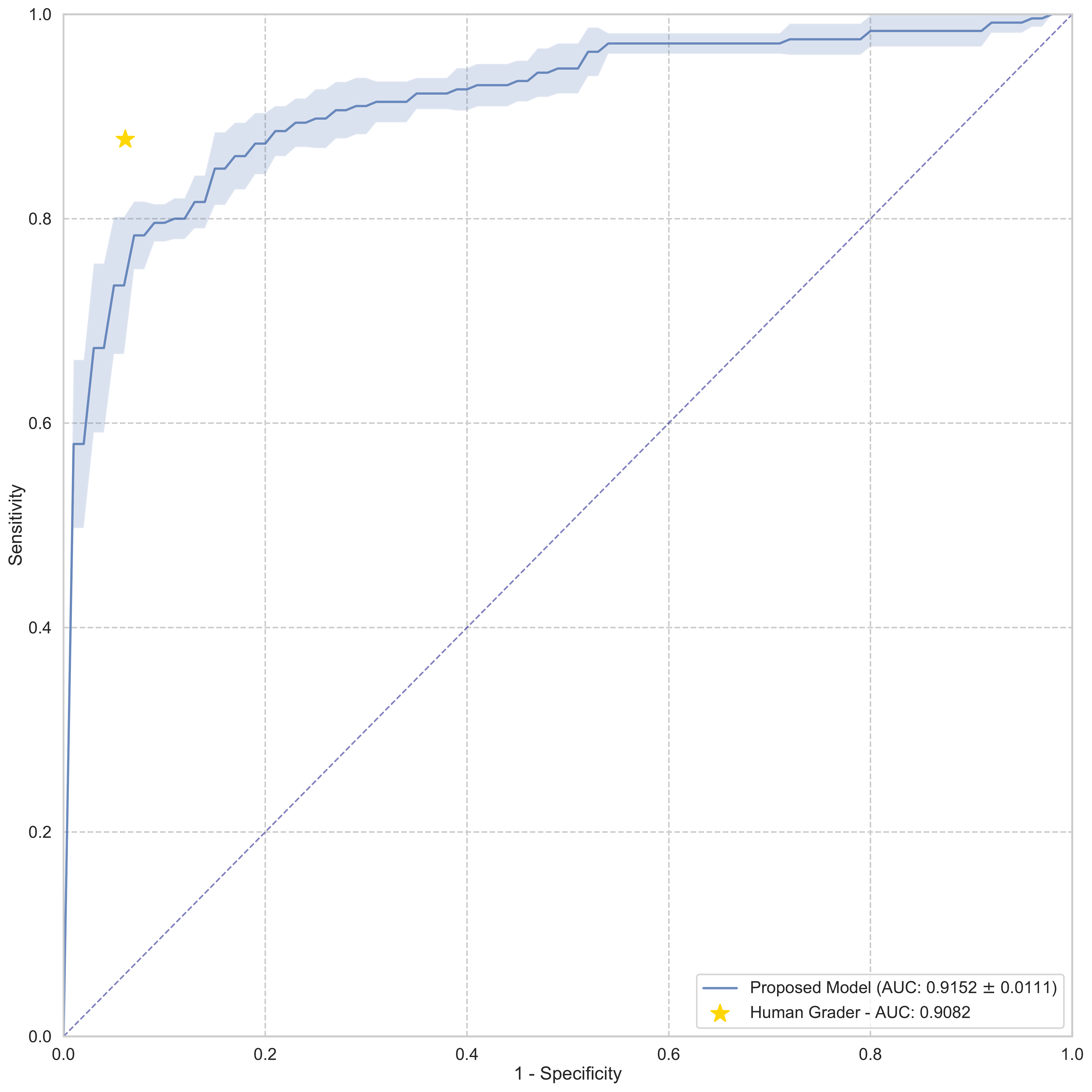}
\caption{AUROC curve for the proposed model on the subset of Stanford test set that was graded by Glaucoma fellowship trained ophthalmologist, with confidence intervals obtained from 5 runs of the model.
To assign a ground truth label, human grader had access to other screening data, including fundus images, OCT RNFL and GCIPL printouts, IOP values, visual field parameters, and also had access to patient history and physical examination data, while the model only had access to the OCT scan cube.}
\label{fig:humangrader-roc-curves}
\end{figure}

On the Stanford test set, our model was able to achieve an AUC value of 0.9080 with a sensitivity value of 0.8591, to differentiate between healthy and normal eyes.
The model was able to achieve an AUC value of 0.8016 with a sensitivity value of 0.7299 on Hong Kong dataset, an AUC value of 0.9428 on the India dataset with sensitivity of 0.9312, and an AUC of 0.8729 on the test set from Nepal with a sensitivity of 0.7928.
The complete results of the model are presented in \autoref{tbl:results}.
We also computed the performance of the human grader on a subset of scans from Stanford test set, and the AUC value of human grader was 0.9082.
On the same subset, our proposed model was able to achieve an average AUC value of 0.9152 (see \autoref{fig:humangrader-roc-curves}).
Note that during training, only a subset of the data from the Stanford was used and no data from Hong Kong, India, or Nepal were used during training.
Fine-tuning the model on the external data sources will result in increased accuracy on the external test set.

False predictions were analysed on the Stanford test set, as can be seen in \autoref{tbl:falsepreds}.
Among the 15 false positive cases, age $>80$ years was the only identifiable clinical feature that could be attributed as a possible reason for these cases being identified falsely positive by the algorithm.
Among the 34 cases identified as false negative by the model, 53.8\% were mild glaucoma cases with mean deviation $>-6$.
Cup Disc ratio, disc size, degree of myopia, and myopic features were not identified as reasons for false predictions.
Myopia was not associated with either false positive or false negative predictions.
But these numbers were too small to make any definitive conclusions.

We also analyzed the performance of the model separately for each myopia severity level.
We defined severity of myopia by slightly modifying the Blue Mountain Eye Study (BMES) \cite{mitchell1999relationship}.
We modified the BMES category of moderate to severe myopia ($> -3 D$) by further subdividing it into mild myopia (up to $-3 D$) moderate myopia ($-3 D$ up to $-6 D$) and severe myopia ($> -6 D$), using cutoffs established in the Beijing Eye Study \cite{xu2007high}.
As can be seen in \autoref{tbl:severityresults}, the model was able to achieve a maximum F1 score of 0.9670 on severe myopia cases, and maximum accuracy of 0.9947 on severe myopia cases.
Model was also able to achieve a maximum F1 score of 0.9057 and AUC of 0.9792 on moderate myopia cases.
Performance on the mild myopia cases were lower than the severe and moderate myopia cases.
The model achieved F1 score of 0.8364, and AUC of 0.8768 on mild myopia cases.

\subsection{\textsc{DiagFind} Experiment}

\begin{figure}[h!]
\begin{subfigure}{.5\textwidth}
  \centering
  \includegraphics[width=.95\linewidth]{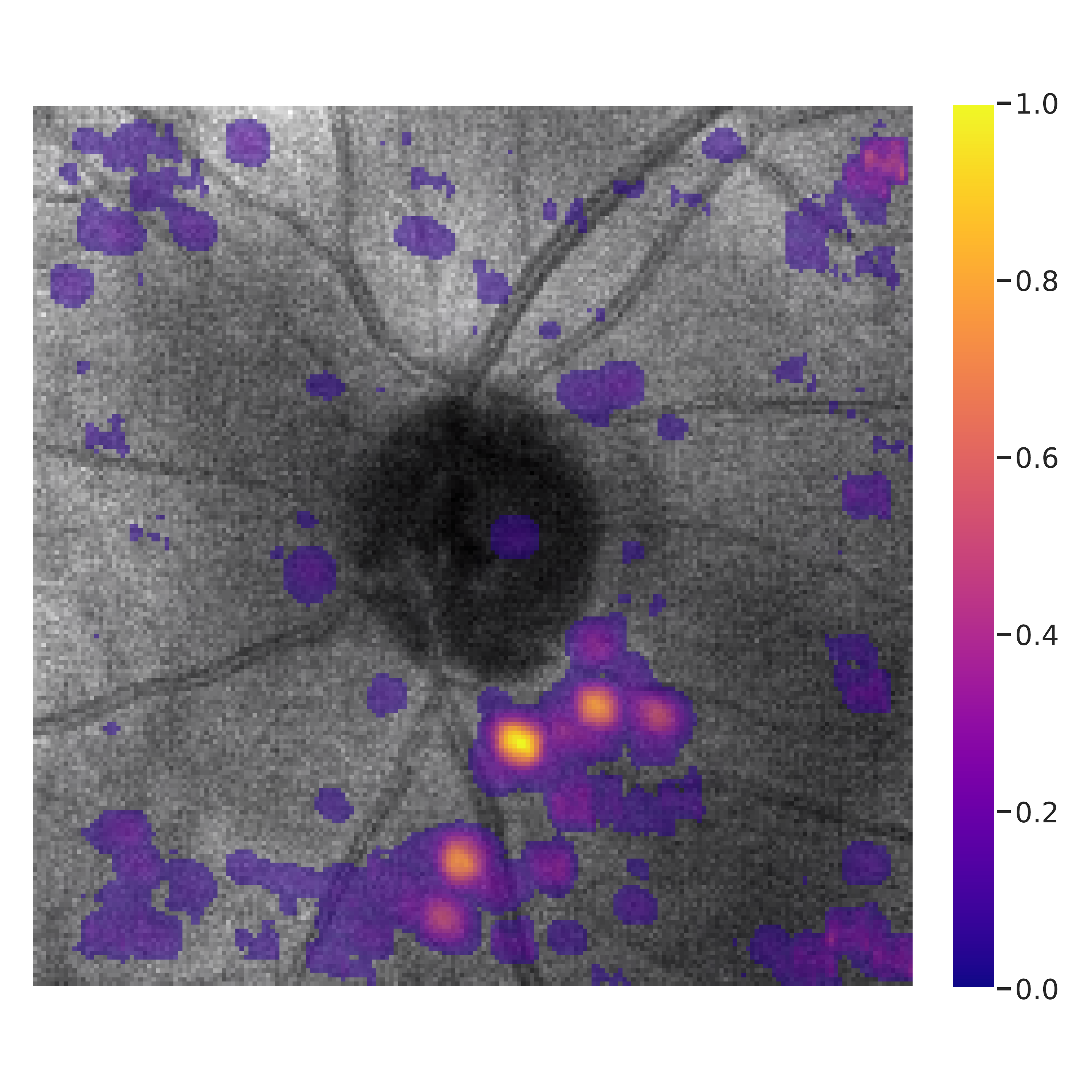}
  \caption{}
  \label{fig:trueglaucomatop}
\end{subfigure}%
\begin{subfigure}{.5\textwidth}
  \centering
\includegraphics[width=.95\linewidth]{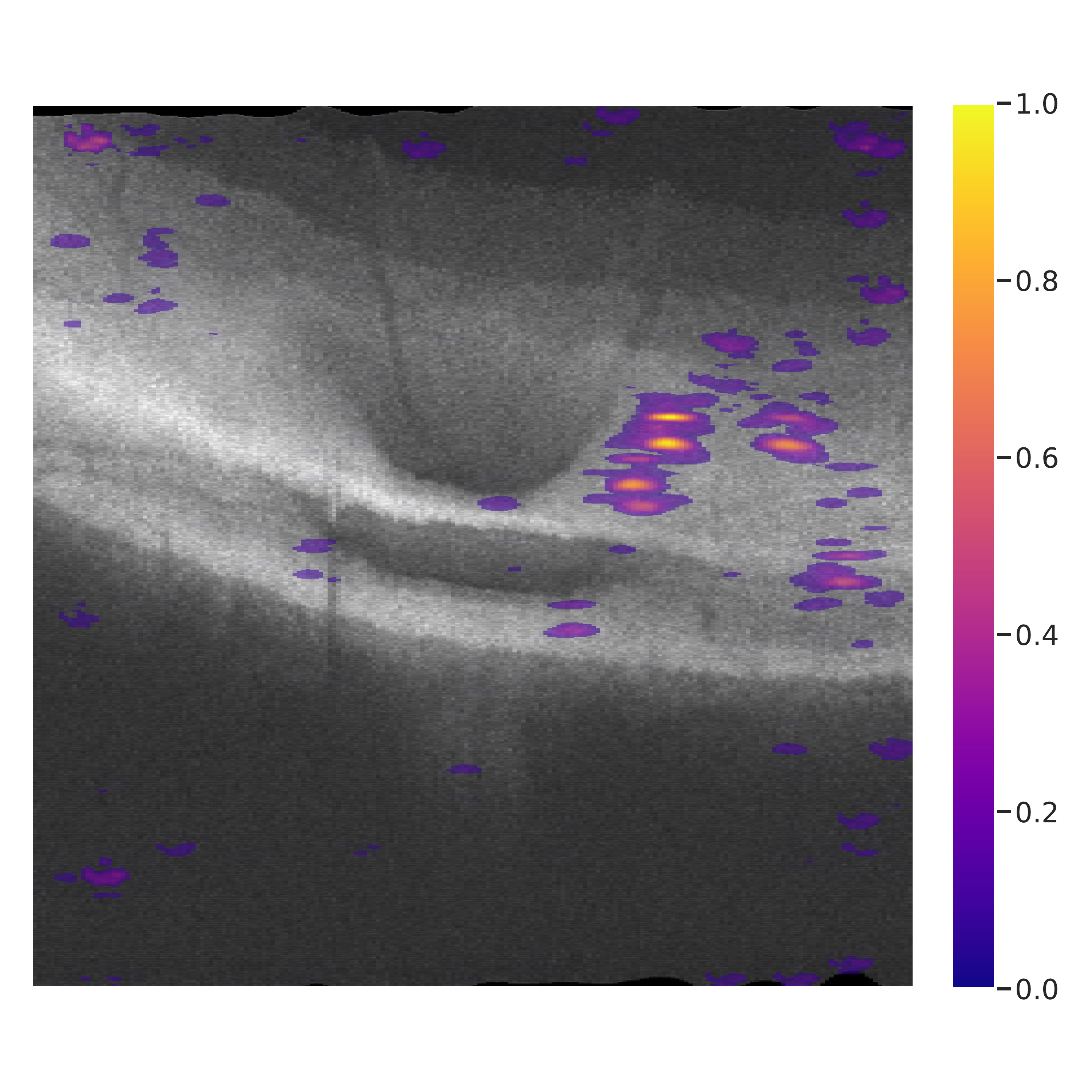}
  \caption{}
  \label{fig:trueglaucomaside}
\end{subfigure}
\\
\begin{subfigure}{.5\textwidth}
  \centering
  \includegraphics[width=.95\linewidth]{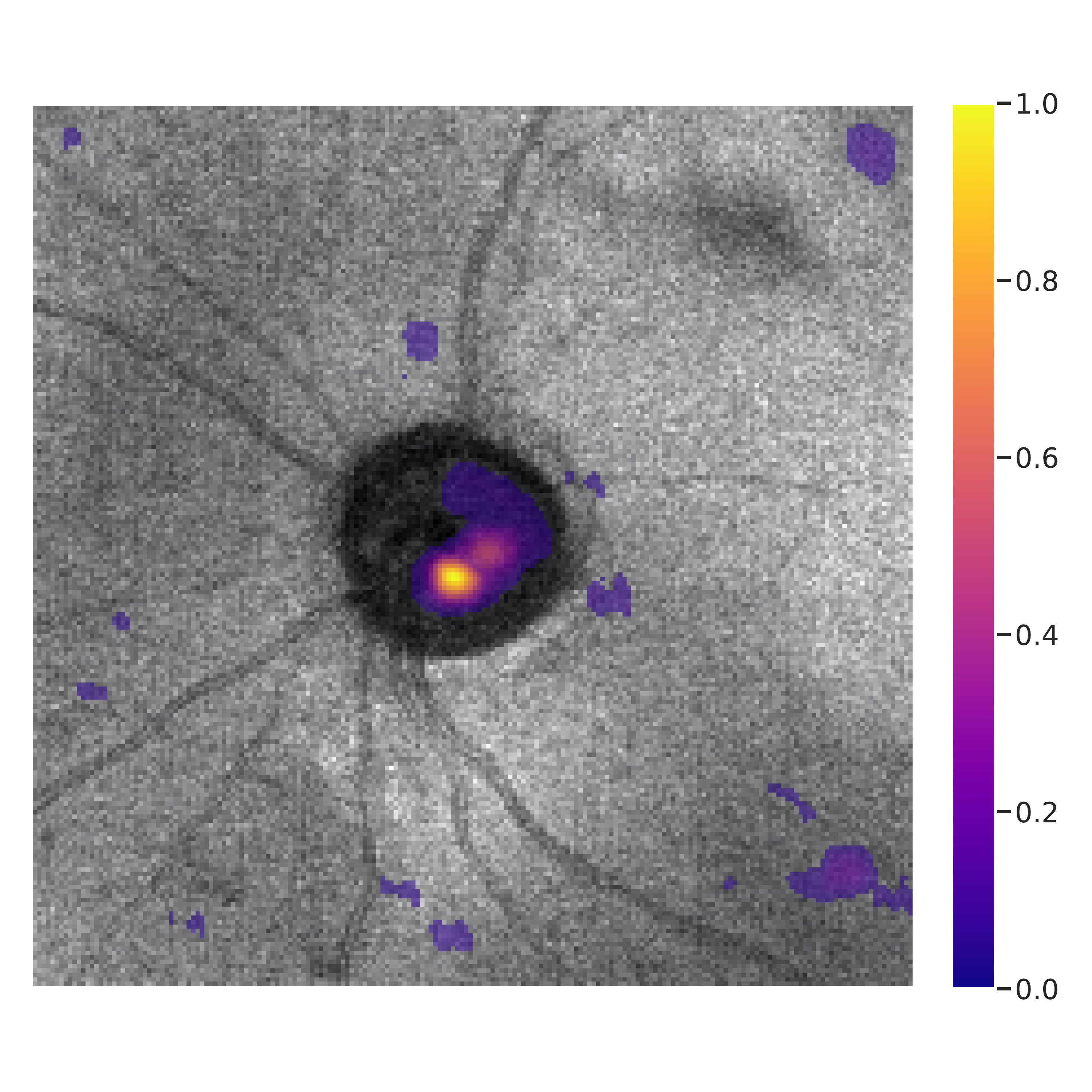}
  \caption{}
  \label{fig:truenormaltop}
\end{subfigure}%
\begin{subfigure}{.5\textwidth}
  \centering
\includegraphics[width=.95\linewidth]{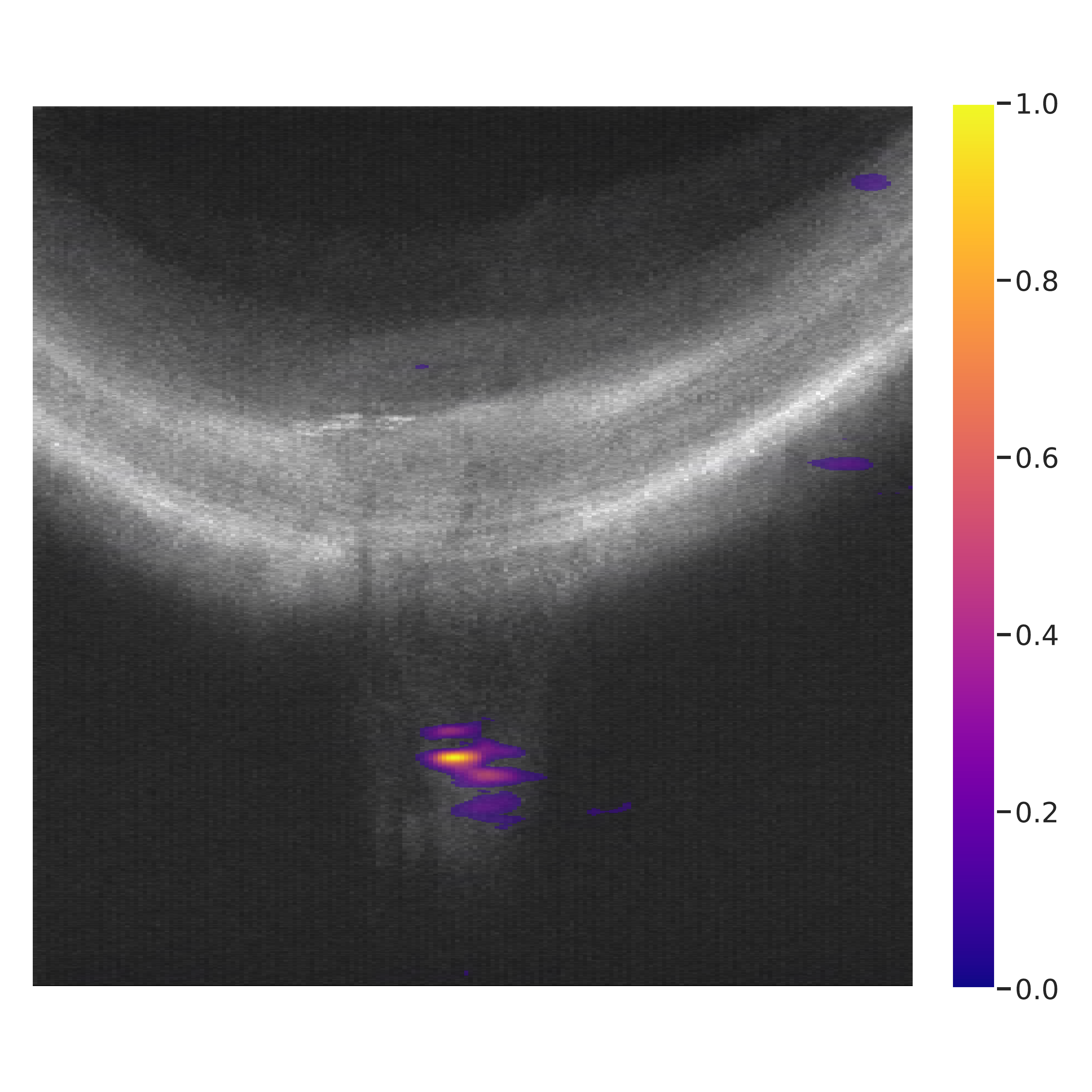}
  \caption{}
  \label{fig:truenormalside}
\end{subfigure}
\caption{Saliency visualizations for two cases from the Stanford Test set.
\textbf{(a)} Top, and \textbf{(b)} Side side view of saliency visualizations of a correctly classified normal eye.
\textbf{(c)} Top, and \textbf{(d)} Side view of saliency visualizations of a correctly classified glaucomatous eye.
As can be seen, in most of the cases, a highlight in the lamina cribrosa region is mostly correlated with \textit{True Glaucoma} prediction, while for cases with \textit{True Normal} prediction, the retinal layer is mostly highlighted.
Saliency visualization have been obtained with respect to the predicted class.
Regions with higher value are more salient for the model in making the final prediction.}
\label{fig:saliency}
\end{figure}

Saliency visualizations show that in most of the cases in which the model makes a \textit{True Glaucoma} prediction, the Lamina Cribrosa is highlighted (see \autoref{fig:trueglaucomatop} and \autoref{fig:trueglaucomaside}).
Out of the 156 cases predicted as \textit{True Glaucoma} by the model on the Stanford test set, all the cases had Lamina Cribrosa highlighted on the saliency visualizations, with or without retina highlighting.
However, when the prediction is \textit{True Normal}, superficial retina is highlighted in a high number of cases (see \autoref{fig:truenormaltop} and \autoref{fig:truenormalside}).
Out of the 92 cases predicted as \textit{True Normal}, (67.3\%) had superficial retina highlighting

Based on these observations, we utilized the \autoref{alg:diagfind} and re-trained the model using additional random cropping data augmentation.
In this data augmentation, we found a heuristic to select the subset of the scan that would contain the lamina cribrosa with a high probability.
During training, the data augmentation would randomly select a subset of the scan cube, and set all the values outside the selected cube as zero.
Note that in this data augmentation, cube sampling was implemented in a way that the heuristically-identified scan cube would be selected with a higher priority compared to other plausible subset cubes.
While the initial model (without random cropping data augmentation) was able to achieve an AUC value of 0.4117 on the manually cropped test set, the model trained on the same data with the same hyperparameters, with the addition of the random cropping data augmentation increased the AUC value to 0.6867.

We also tried this experiment by utilizing more training data from each of the external test sets, in addition to the training set from Stanford.
20\% of the cases from each external test set were randomly selected.
This resulted in 322 additional \textit{True Normal} scans and 474 additional \textit{True Glaucoma} scans.
Using these additional data, and re-training using the procedure described in the \autoref{alg:diagfind}, the AUC on the cropped scans increased to 0.7700, which is a substantial relative increase.

\begin{table}[h!]
\centering
\caption{Results of the proposed model on the Stanford and external test sets.
Mean and standard deviation were computed over five runs of the model with different seeds but the same values of hyperparameters.}
\label{tbl:results}
\begin{tabular}{@{}lcccc@{}} \toprule
\textbf{Dataset} & \textbf{AUC} & \textbf{Sensitivity} & \textbf{Specificity} & \textbf{F1 Score} \\ \midrule
    \textbf{Stanford} & $0.9080~(\pm 0.0051)$ & $0.8591~(\pm 0.0474)$ & $0.7776~(\pm 0.0827)$ & $0.8687~(\pm 0.0108)$\\
\textbf{Hong Kong} & $0.8016~(\pm 0.0133)$ & $0.7299~(\pm 0.0506)$ & $0.7306~(\pm 0.0998)$ & $0.7613~(\pm 0.0069)$ \\
\textbf{India} & $0.9428~(\pm 0.0090)$ & $0.9312~(\pm 0.0439)$ & $0.7098~(\pm 0.1600)$ & $0.9071~(\pm 0.0073)$ \\
\textbf{Nepal} & $0.8729~(\pm 0.0161)$ & $0.7928~(\pm 0.0891)$ & $0.7870~(\pm 0.1055)$ & $0.8034~(\pm 0.0225)$ \\ \bottomrule
\end{tabular}
\end{table}

\begin{figure}[h!]
\centering
\includegraphics[width=\linewidth]{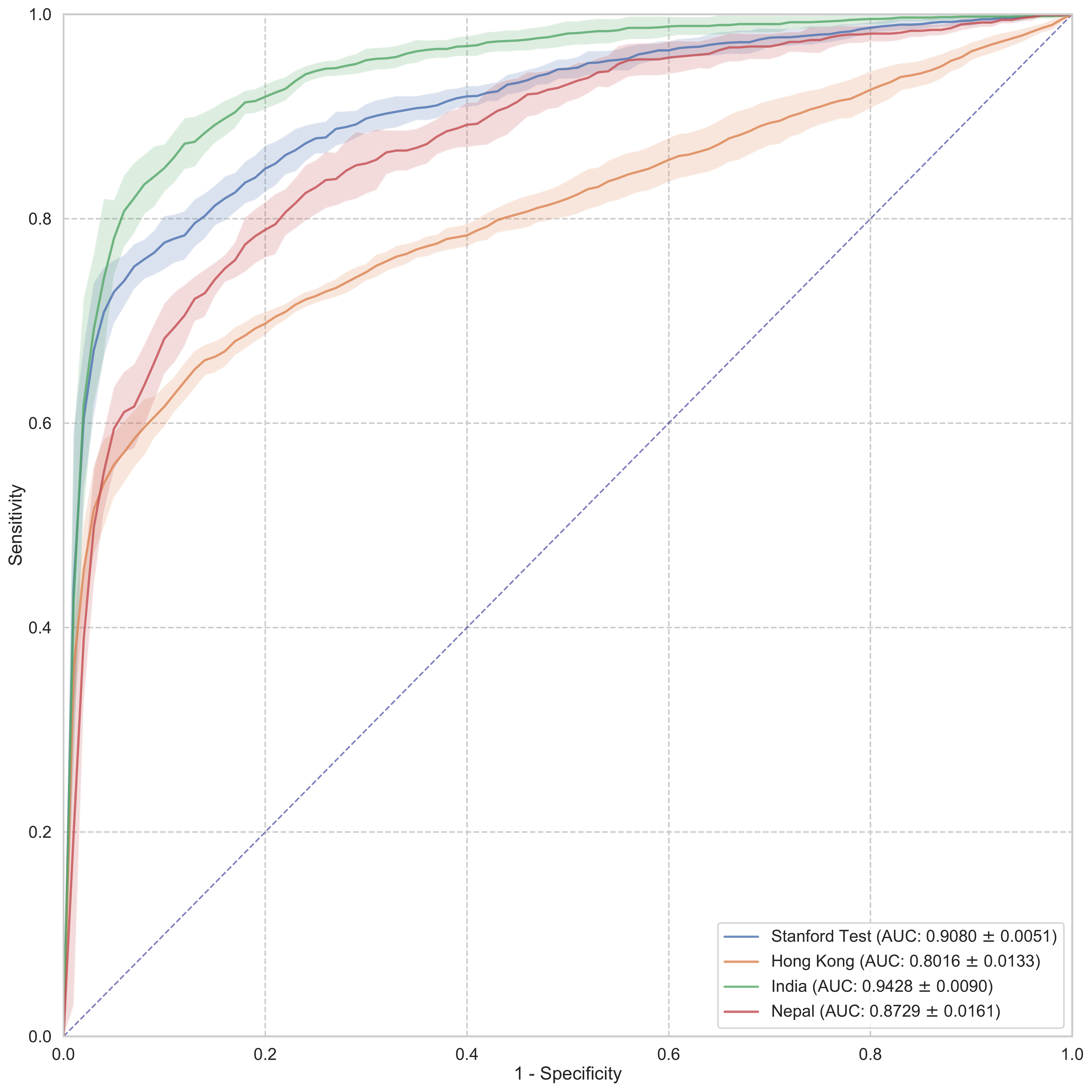}
\caption{AUROC curves for all the test sets, with confidence intervals obtained from 5 runs of the model.}
\label{fig:roc-curves}
\end{figure}

\begin{table}[h!]
\centering
\caption{Results of the proposed model on the Stanford test set for each myopia severity level.
Mean and standard deviation values have been computed over 5 runs.}
\label{tbl:severityresults}
\begin{adjustbox}{width=\textwidth,center}
\begin{tabular}{@{}lccccc@{}} \toprule
\textbf{Myopia Severity} & \textbf{Number of scans} & \textbf{AUC} & \textbf{Sensitivity} & \textbf{Specificity} & \textbf{F1 Score} \\ \midrule
\textbf{Mild} & 166 & 0.9174 ($\pm 0.0250$) & 0.8937 ($\pm 0.0390$) & 0.6909 ($\pm 0.0750$) & 0.8841 ($\pm 0.0370$) \\
\textbf{Moderate} & 52 & 0.9646 ($\pm 0.0292$) & 0.9143 ($\pm 0.0649$) & 0.8917 ($\pm 0.0697$) & 0.9248 ($\pm 0.0487$) \\
\textbf{Severe} & 51 & 0.9872 ($\pm 0.0117$) & 0.9447 ($\pm 0.0242$) & 0.9000 ($\pm 0.1369$) & 0.9729 ($\pm 0.0080$) \\ \bottomrule
\end{tabular}
\end{adjustbox}
\end{table}

\begin{table}[h!]
\centering
\caption{Results of the proposed model trained with the \textsc{DiagFind} algorithm, on the cropped scans from Stanford test set for each myopia severity level.
Note that number of cropped scans with myopia severity information that have severe and moderate levels of myopia is very small.}
\label{tbl:croppedmyopiaseverity}
\begin{tabular}{@{}lccccc@{}} \toprule
\textbf{Myopia Severity} & \textbf{Number of scans} & \textbf{AUC} & \textbf{Sensitivity} & \textbf{Specificity} & \textbf{F1 Score} \\ \midrule
\textbf{Mild} & 24 & 0.7714 & 0.7143 & 0.50000 & 0.6897 \\
\textbf{Moderate} & 7 & 0.7500 & 0.7500 & 0.6667 & 0.7500 \\
\textbf{Severe} & 4 & 1.0000 & 1.000 & 1.0000 & 1.0000 \\ \bottomrule
\end{tabular}
\end{table}

\begin{table}[h!]
\centering
\caption{Results of the proposed model on the Stanford test set for each Glaucoma severity level, for scans where we have Glaucoma severity information.
Mean and standard deviation values have been computed over 5 runs.}
\label{tbl:glaucomaseverity}
\begin{tabular}{@{}lcc@{}} \toprule
\textbf{Glaucoma Severity} & \textbf{Number of scans} & \textbf{Recall} \\ \midrule
\textbf{Mild} & 225 & 0.8373 ($\pm 0.0824$) \\
\textbf{Moderate} & 70 & 0.9200 ($\pm 0.0217$) \\
\textbf{Severe} & 66 & 0.9848 ($\pm 0.0151$) \\ \bottomrule
\end{tabular}
\end{table}

\begin{table}[h!]
\centering
\caption{Observed causes of false predictions of \textit{True Glaucoma} versus \textit{True Normal} on the Stanford test set.
Cup Disc ratio, disc size, and myopic features were not identified as reasons for false predictions.}
\label{tbl:falsepreds}
\begin{tabular}{@{}llc@{}} \toprule
\multicolumn{2}{@{}l}{\textbf{False Predictions}} & \textbf{Number of eyes} \\ \midrule
\multicolumn{2}{@{}l}{\textbf{False Positives}} & 15 \\
 & Age $> 80$ & 7 (46.6\%) \\ \midrule

\multicolumn{2}{@{}l}{\textbf{False Negatives}} & 34 \\
 & Mild Glaucoma MD $>-6$ & 53.8\% \\
\bottomrule
\end{tabular}
\end{table}

\begin{figure}[h!]
\begin{subfigure}{.5\textwidth}
  \centering
  \includegraphics[width=.95\linewidth]{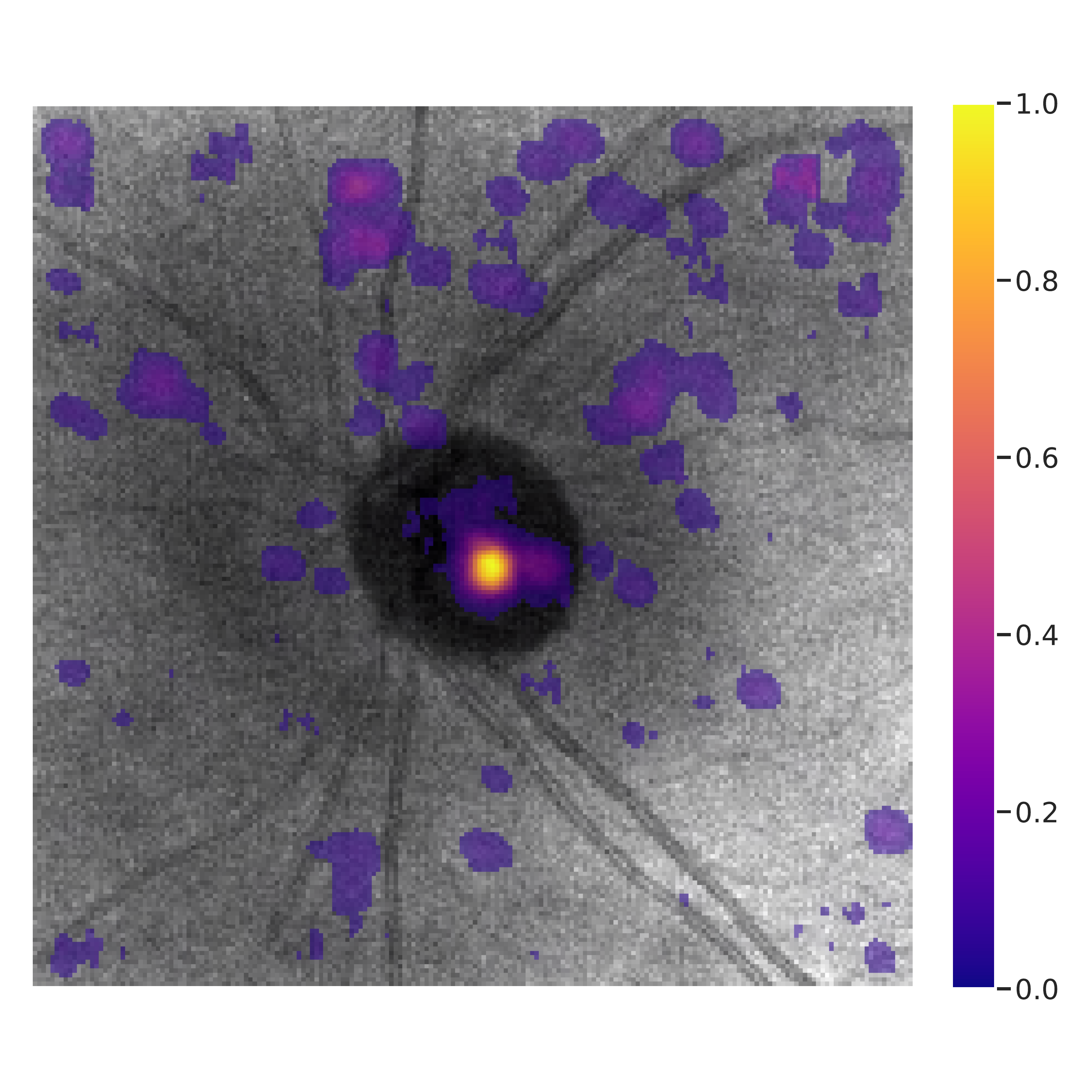}
  \caption{}
  \label{fig:falseposhkatop}
\end{subfigure}%
\begin{subfigure}{.5\textwidth}
  \centering
\includegraphics[width=.95\linewidth]{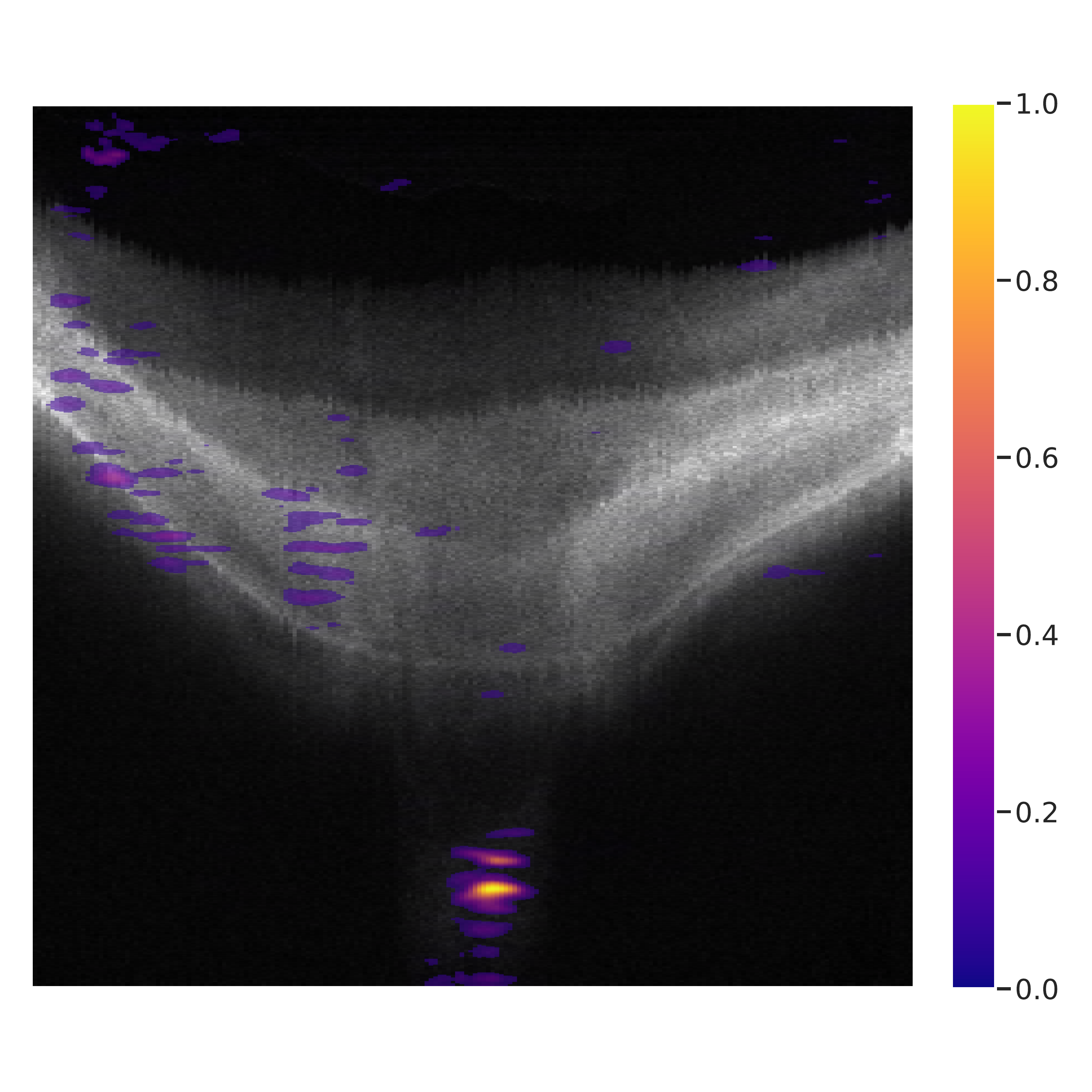}
  \caption{}
  \label{fig:falseposhkside}
\end{subfigure}
\\
\begin{subfigure}{.5\textwidth}
  \centering
  \includegraphics[width=.95\linewidth]{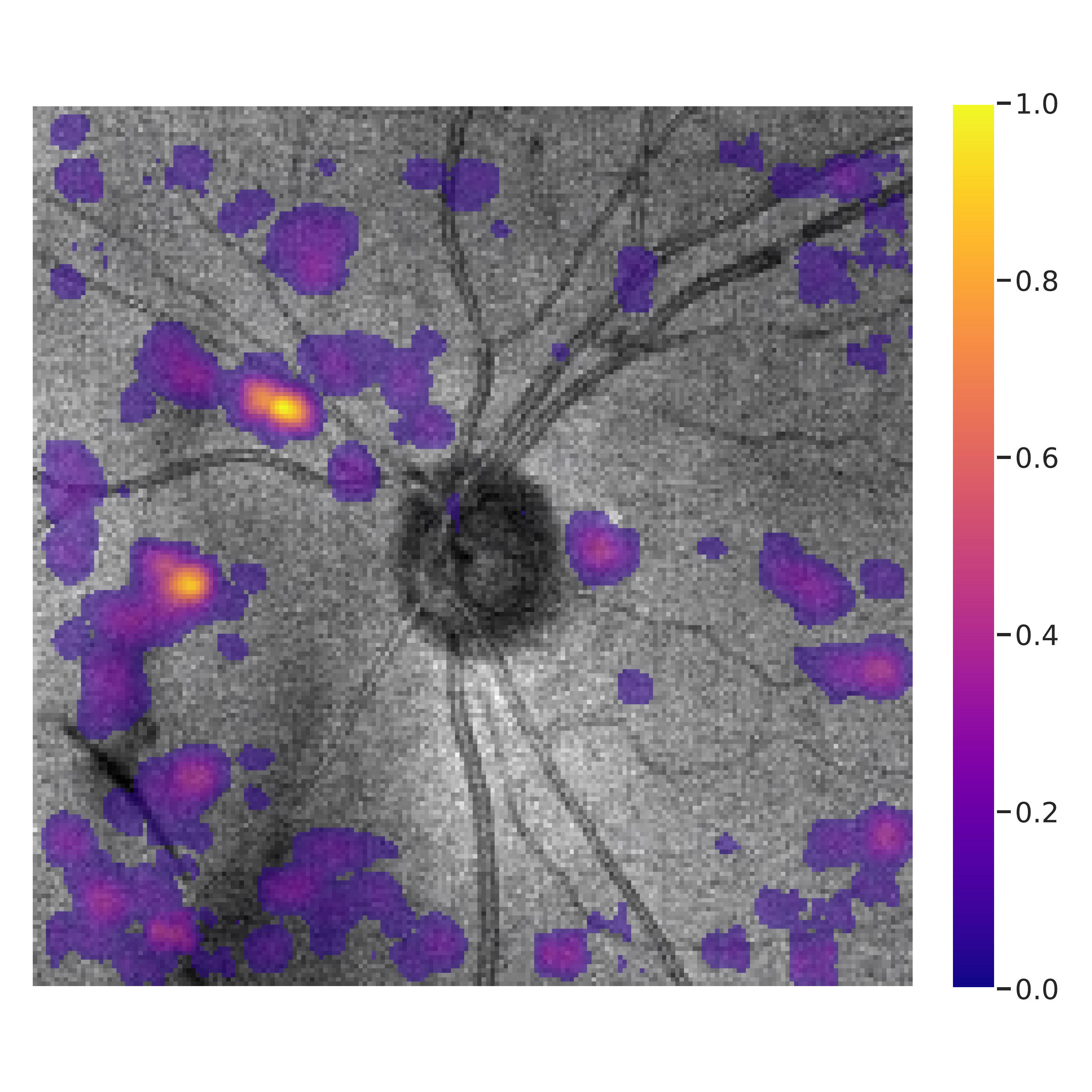}
  \caption{}
  \label{fig:falsenegindiatop}
\end{subfigure}%
\begin{subfigure}{.5\textwidth}
  \centering
\includegraphics[width=.95\linewidth]{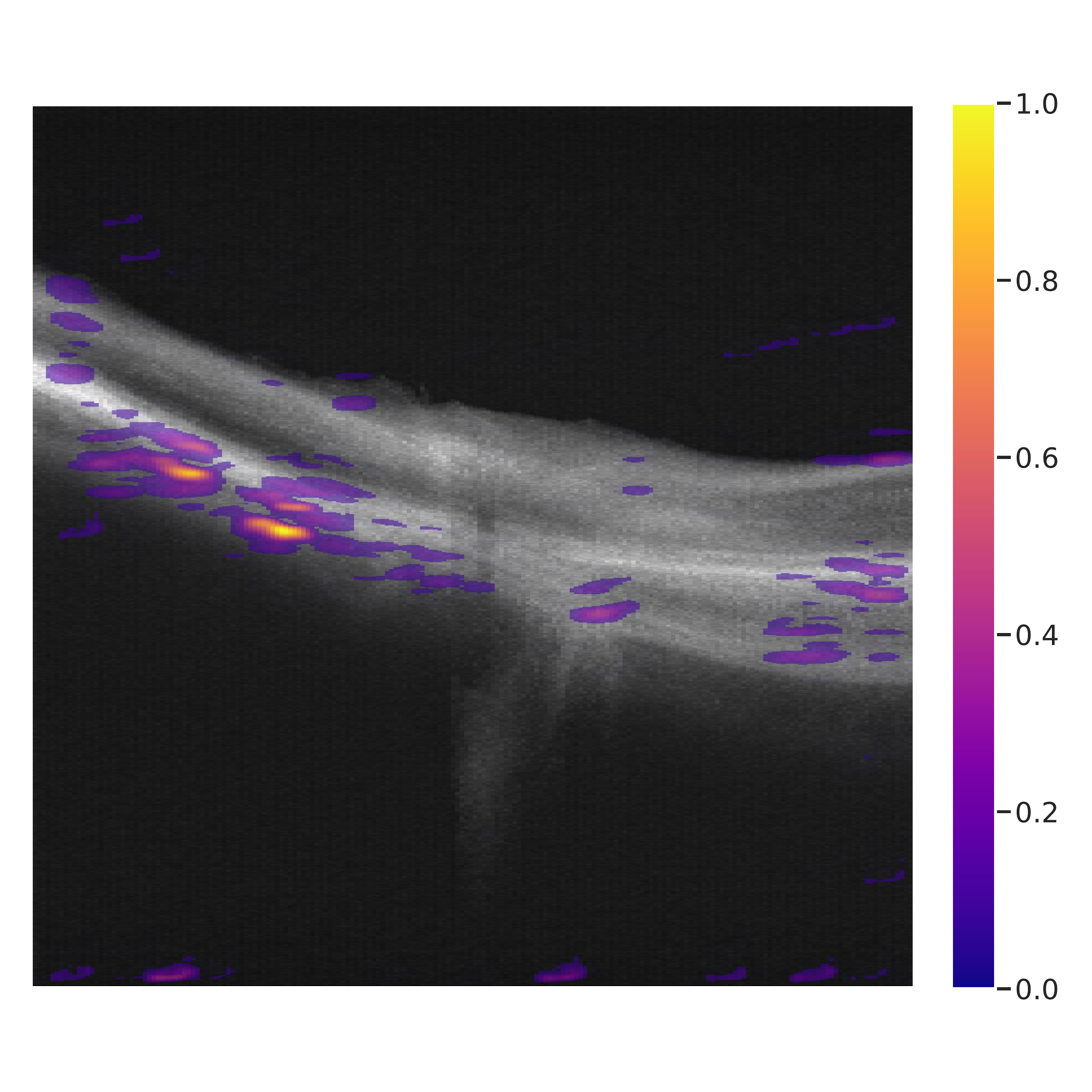}
  \caption{}
  \label{fig:falsenegindiaside}
\end{subfigure}
\caption{Saliency visualizations for two cases from the Stanford test set with wrong predictions.
\textbf{(a)} Top, and \textbf{(b)} Side side view of saliency visualizations of a false positive case, where Lamina Cribrosa is highlighted, even though the case has \textit{True Normal} ground truth label.
\textbf{(c)} Top, and \textbf{(d)} Side view of saliency visualizations of a false negative case, where the retina is highlighted despite the case having \textit{True Glaucoma} ground truth.
Saliency visualization have been obtained with respect to the predicted class.
Regions with higher value are more salient for the model in making the final prediction.}
\label{fig:saliencywrongresults}
\end{figure}

\section{Discussion}

In this study we developed and validated a 3D deep learning system using real world raw OCT optic nerve head volumes to detect glaucomatous optic neuropathy from normals.
The labeled ground truth of glaucoma was assessed by reviewing fundus photos, OCT RNFL and macula results, visual field results, and IOP and treatment data over several visits to make sure there was no doubt glaucoma was present.
Since the definition of glaucoma is very important when training an algorithm, we realized there is a limitation in diagnosing glaucoma with just the OCT red/yellow/green printout, and thus the qualitative RNFL and GCIPL thickness and deviation maps were reviewed.
We used the criteria defined in \autoref{suptbl:criteria} during data labeling.
While many of the recent studies (\textit{e.g.}~\cite{ran2019detection}) define structural changes in glaucoma based on OCT RNFL thickness and/or deviation maps alone, we think the additional multimodal test results allow for training an algorithm on a wider variation in the population instead of narrowing the inclusion criteria.

In our study, the machine learning system performed with an AUC of 0.9080 ($\pm 0.0051$) to differentiate between healthy and definite glaucomatous eyes of all ranges from early perimetric to late perimetric glaucoma.
Our performance with external testing generalized across multinational datasets where there are differing patient populations with varying disease severities.
The performance was also very good on the external test set from India, which had an AUC value of 0.9428 ($\pm 0.0090$).
We hypothesize that this is because there was a significantly higher percentage of eyes with severe disease in this dataset compared to other external datasets (Supplementary \autoref{suptbl:additional}).
These cases would likely be easier to differentiate from normal cases.
The performance was reduced on the dataset from Hong Kong with an AUC of 0.8016 ($\pm 0.0133$).
This can be explained due to the differences in their labeling criteria which defined structural changes in glaucoma based on the RNFL thickness and/or deviation maps alone.
Also there was significant differences in the refractive error between the \textit{True Glaucoma} and  \textit{True Normal} cases in the Stanford dataset and the Hong Kong external test set.
Mean refractive error in \textit{True Glaucoma} cases in Hong Kong dataset was $-0.85$ ($\pm 2.57$) versus $-2.16$ ($\pm 4.17$) in the Stanford training dataset.
Mean refractive error in the \textit{True Normal} cases was $-0.51~(\pm 2.15)$ in Hong Kong dataset versus $-2.2~(\pm 4.62)$ in the Stanford training data.
Another reason for the difference in performance on the test set from Hong Kong could be the inclusion of solely gradable images with signal strength $\geq 5$.
We included cases with signal strength $\geq 3$ and excluded images with artifacts which obscured imaging of ONH and the area inside and including the RNFL measurement circle at 3.4 mm from the center of the ONH.
This is because many at times, clinicians are deprived of high quality OCT images for diagnosis and evaluation of glaucoma, due to medial opacity, tear film issues, small pupils, or other limitations.
Our aim was to train the algorithm to be able to identify representations to detect glaucoma even on low quality images, hence replicating real world presentations.
Even though it is recommended to obtain scans of signal strength higher or equal to 6 to facilitate the longitudinal quantitative progression calculated parameters, qualitative patterns in thickness and deviation maps can still be seen at lower signal strength, and it is suggested that signal strength of $>3$ is acceptable to obtain reproducible scanning images among patients with ocular media opacities \cite{ha2012low}.

With our fourth external dataset from Nepal, the model performed with an AUC of 0.8729 ($\pm 0.0161$).
Possible explanations for the difference in performance could be due to the differences in the dataset.
The mean age of the subjects were significantly lower in this dataset.
There was a statistically significant difference in female to male ratio in the dataset from Nepal.
The percentage of eyes with severe myopia in the \textit{True Glaucoma} and \textit{True Normal} subsets were lower compared to the data from Stanford.
The mean refractive error was significantly lower in this dataset.
Another possible reason for the differences in performance across the external datasets could be possible inter- and intra-grader variability in labeling of cases based on the criteria.
Given the variance in the age, refractive error, glaucoma severity, demographic features such as ethnicity, gender distribution, and differences in labeling criteria across datasets, we think that the performance of the algorithm imply potential for utilization in real world clinical scenarios. % Amazing line!
We asked a Glaucoma fellowship trained ophthalmologist to grade a subset of cases from the Stanford test set.
The human grader was able to achieve an AUC of 0.9082, while the proposed model was able to achieve an AUC value of 0.9152 ($\pm 0.0111$) on the same subset.
From 5 runs of our model, the worst p-value obtained was 0.3670, which shows that the difference in the performance of the proposed model and the human grader is not statistically significant~\cite{draelos_2020,pROC,delong1988comparing}.
Note that the human grader had access  to other screening data, including fundus images, OCT RNFL and GCIPL printouts, IOP values, visual field parameters, and also had access to patient history and physical examination data.

A novel output of our model is its ability to detect glaucoma across different ranges of myopia (\autoref{tbl:severityresults}).
The model was able to achieve an accuracy of 0.9947 on severe myopia cases, AUC of 0.9792 on moderate myopia, and accuracy of 0.8768 on mild myopia cases.
It is known that diagnosing glaucoma in the setting of myopia is a common challenge due to alteration of the appearance of the optic nerve and OCT.
Myopic refractive error impacts RNFL and macular thickness measurements due to stretching and thinning of these layers and due to increased axial length and optical projection artifact of the scanning area \cite{mwanza2012rates}.
This often results in many false positive diagnoses, also known as ``Red Disease".
Using the entire cube and highlighting the lamina cribrosa may help researchers study this LC region more closely in myopes when trying to differentiate glaucoma from normal.
The difference in the performance in the myopia subsets compared to the total dataset could be due to the fewer number of cases in each subgroup (\autoref{tbl:severityresults}).

What was most interesting from our model were the saliency maps of the regions in the scan where the model attends to make a prediction.
Normally, we expected the RNFL to be a majority of the differentiation of true glaucoma from normal, but in many cases, the lamina cribrosa was just as important, or sometimes more important since the RNFL can be thinned for other reasons such as myopia.
Given that clinicians do not routinely review every single slice of the cube, and because current OCT RNFL and ONH printouts do not provide any diagnostic information based on LC, we were excited to discover that, by training a model on every single slice, saliency visualization highlighted the lamina cribrosa region along with exiting nerve fibers posterior to LC, and in most cases are correlated with \textit{True Glaucoma} prediction.
For cases with \textit{True Normal} prediction, the areas on superficial retina were mostly highlighted in saliency visualizations.
This corresponds with clinical practice, whereby when an OCT RNFL is all normal (all RNFL quadrants colored green or white), then likely it has a very high negative predictive value for glaucoma.

Our assessment of false predictions by the 3D deep neural model showed no correlation with myopia, despite the fact that myopia is one of the most common reason for misdiagnosis of glaucoma in clinical presentations \cite{mwanza2012rates}.
This suggests that by training the model on all scans including high myopes and low signal strength ones as long as there were no data loss artifacts, could provide enough training examples within the volumes of slices to avoid myopia affecting the result.
Following the identification of LC as area of interest on our saliency maps in \textit{True Glaucoma}, we were motivated to analyze the diagnostic information content of the lamina cribrosa area by using the \textsc{DiagFind} algorithm and training and evaluating the model on scans cropped to only contain the optic nerve head.
With random cropping data augmentation, we achieved an AUC of 0.6867 on cropped scans.
When we used additional training data and experimented on the cropped scans, the performance increased to 0.7700 which is a measurable in absolute difference in performance, even though the difference is not statistically significant ($p = 0.0706$)~\cite{draelos_2020,pROC,delong1988comparing}.
Because peripheral LC or regions posterior to large blood vessels typically remain difficult for OCT image interpretation without enhanced depth imaging, our algorithm's ability to utilise the diagnostic information at the LC from the conventional scans (without enhanced depth imaging) can be a turning point in Glaucoma diagnosis, especially in cases of high myopia and severe glaucoma cases.
In severe degrees of myopia, retinal parameters are less useful in evaluation of glaucoma~\cite{mwanza2012rates}.
We evaluated the performance of the algorithm across different severity levels of myopia cases on the cropped scans (see \autoref{tbl:croppedmyopiaseverity}).
The algorithm performed with an AUC of 0.7714, 0.7500, and 1.000 on mild, moderate, and severe myopia cases, respectively.
Even though more validation is required due to the small data distribution, the results point towards possibility of utilization of diagnostic information at the lamina cribrosa level in different degrees of myopia in glaucoma diagnosis, especially as an alternative to unreliable RNFL parameters in severe myopia.
In severe glaucoma cases, RNFL thickness levels off, falling below $50~\mu m$ but almost never below $40~\mu m$ for the Cirrus machine, due to the assumed presence of residual glial or non-neural tissue including blood vessels, and hence makes RNFL measurement less clinically useful for identifying progression~\cite{hood2007framework}.
We evaluated the performance of our algorithm across severity of glaucoma (see \autoref{tbl:glaucomaseverity}).
The model performed was able to achieve recall values of 0.8373, 0.9200, and 0.9848, on average, on mild, moderate, and severe glaucoma, respectively.
Even though this needs more analysis with a larger distribution of glaucoma severity based datasets, the results show possibility of utilizing information alternative to RNFL thickness in prognosis of severe glaucoma, overcoming the floor effect.

Recently Maetschke \textit{et al.}~\cite{maetschke2019feature} employed 3D convolutional neural networks to classify eyes as healthy or glaucomatous directly from raw, unsegmented OCT volumes (1110 scans) of the optic nerve head obtained using Cirrus SD-OCT scanner (Carl Zeiss Meditec Inc., Dublin, CA, USA) and achieved a substantially high AUC of 0.94 against logistic regression, which was found to be the best performing classical machine learning technique with an AUC of 0.89.
In their study, glaucomatous eyes were defined as those with glaucomatous visual field defects alone and was not based on any structural parameters.
This work used a convolutional neural network for the task of glaucoma classification, however, the architecture used for the neural network was different from the architecture of the proposed model.
Another difference from our study was that they included scans with signal strength $\geq 7$.
Despite the differences in definition and inclusion criteria, it is interesting to note that our saliency maps had similar findings.
Similar to our study, for healthy eyes, the network in \cite{maetschke2019feature} tends to focus on a section across all layers and ignores the optic cup/rim and the lamina cribrosa.
In contrast, for glaucomatous eyes, the optic disc cupping, neuroretinal rims, as well as the lamina cribrosa and its surrounding regions were highlighted.
The strength of our study compared to \cite{maetschke2019feature} is that we included more information about our training population and had multiple external datasets for validation.
Additionally, we devised a new experiment to understand how much diagnostic information is contained in the lamina cribrosa and optic nerve head area.

In the recent study by Ran \textit{et al.}~\cite{ran2019detection}, the 3D deep learning system had an AUC of 0.969.
The study showed good performance with external test set from United States with an AUC of 0.893.
Similar to our study, the heatmaps generated in their study showed neuro-retinal rim and areas covering the lamina cribrosa to be highlighted in detection of glaucomatous optic neuropathy.
Apart from this, the retinal nerve layer and choroid were also potentially found be related to detection of glaucomatous optic neuropathy in their study.
The difference in their study from ours was in the definitions used for glaucoma and inclusion of images with signal strength $\geq 5$. They defined glaucomatous structural defect based on OCT RNFL thickness and deviation maps.

While it is unclear about the distribution or inclusion of different degrees of myopia in their study, our cohort had 8.8 percentage of total eyes with severe myopia ($\geq-6$) in our \textit{True Glaucoma} subset and 4.09 percentage of total eyes with severe myopia in the \textit{True Normal} subset in the Stanford data.
Another difference was the distribution of ethnicity in their training set which consisted exclusively of Chinese Asian eyes, while our training, validation, and test data from Stanford included subjects of Caucasian, Asian (which included Chinese Asians, Non-Chinese Asians, and Indians), African American, and Hispanic origin.

The major differences between the recent studies \cite{maetschke2019feature,ran2019detection} and ours was the diversity in the ethnicity of the datasets used for training of the model, inclusion of high refractive errors in both glaucoma and normal cases for training, and inclusion of eyes with lower signal strength, hence representing the real world clinical presentations.
Our work used external test datasets from India, Hong Kong, and Nepal, while similar works (\textit{e.g.} \cite{ran2019detection}) did not have similar variety in the external tests sets used.

Our study has several strengths.
Multiple international datasets provide diversity in our database for evaluation purposes, which is rare to have for glaucoma datasets.
We had images from patients of different ethnicities, including Caucasian, Asian (which included Chinese Asians, Non-Chinese Asians, and Indians), African American, Hispanic, and of Indian origin.
The performance of our model was promising across multiple geographies and ethnicities to distinguish glaucoma from normal.
Another significant strength of our method was that our main training dataset was not cleaned for this experiment to more closely follow the challenges that are faced in real world clinical settings.
While most studies have strict exclusion criteria based on axial length, disc sizes, and high myopia, our cohort included all ranges of myopia, disc sizes, and axial lengths, reflecting real world presentations.
One other major highlight of our study was the criteria used to classify cases as \textit{True Glaucoma} versus \textit{True Normal} in the training and validation dataset, which included both multimodal longitudinal structural and functional evaluations.
This closely replicates real world clinical settings where multimodal longitudinal evaluation is used to arrive at the diagnosis.
Additionally, using the \textsc{DiagFind} algorithm, we were able to show that a new region in the OCT scan, that is mostly ignored by ophthalmologists for detection of glaucoma, indeed contains useful diagnostic information that can serve as an additional signal in the glaucoma diagnosis.
Apart from this, our experiment with cropped scans had  encouraging results for using optic nerve head region with focus on lamina cribrosa in diagnosis and prognosis of the disease, especially in high myopia and severe glaucoma, where conventional RNFL parameters have limitations.

On the other hand, our study has a few drawbacks.
We did not include ``Suspect'' cases in our datasets.
This was mainly because of the difficulty in obtaining consensus for glaucoma suspect definition among experts.
We are now working on a separate dataset and are trying to achieve consensus among multiple glaucoma experts to classify high- and low-risk suspect or referral cases.
Additionally, we have not included ``Preperimetric'' glaucoma in the training due to the unavailability of adequate number of cases in the subset.

Even though we have not excluded any cases based on disc sizes or presence of myopic tilted discs in our datasets, and have included cases with low signal strength, we have not looked into the performance of our model across subsets.

Going forward, we plan to develop a 3D deep learning algorithm using a wider range of data including high- and low-risk suspect cases that would help in identifying cases which require referral for management by glaucoma specialists.
Secondly, we also plan to evaluate the performance across severity of glaucoma cases and look closely at the patterns in each severity subset by including larger number of cases in each subset.
Finally, we plan to include raw OCT macula cube scans along with optic nerve head scans for better algorithm correspondence.

\section{Conclusion}

Our 3D deep learning model was trained and tested using the largest OCT glaucoma dataset so far from multinational data sources, and has been able to detect glaucoma from raw SD-OCT volumes across severity of myopia and severity of glaucoma.
By using a multimodal definition of glaucoma, we could include more scans from the real world including low signal strength, which are typically excluded from studies.
The saliency visualizations highlighted the lamina cribrosa as an important component in the 3D optic nerve head cube in differentiating glaucoma, which may be useful in high myopes who have thin RNFL.
Based on this information, and using the \textsc{DiagFind} algorithm, we studied the performance of the model in the case that only the optic nerve head crop of the full scan was given to the model.
We observed that our model trained with additional random cropping data augmentation was still able to detect Glaucoma on the cropped scans.

\section*{Author Contributions}

RTC, RZ, EN, and SSM made the conception and design of the work described here and contributed equally to the study.
RTC and SSM contributed to the study protocol.
EN and RZ developed the deep learning system and augmentation strategies with clinical inputs from RTC and SSM.
EN created the code used in this work.
EN and SSM performed the data management, data analysis, data anonymization, and literature search.
EN and RZ managed the model training and evaluation.
SSM contributed to the Stanford (United States) data collection and data annotation, coordinated international data collection, performed the statistical analysis and manual cropping of 3D scans.
RTC and SSM generated the ground truth set and interpreted the results.
DC contributed as Human Grader for validation.
EN and SSM wrote the first draft of the paper.
RTC and RZ contributed to the critical review, revision, editing and provided important intellectual content.
ARR collected and anonymized the Hong Kong data and coordinated ethical approvals for HK data transfer.
SD and MR collected and anonymized the India data.
SST, HLR, and CYC supervised the data collection in Nepal, India, and Hong Kong, respectively.
SN and HLR coordinated ethical approvals for India data transfer and SST coordinated ethical approvals for Nepal data transfer.
CCT supervised the study in Hong Kong.
All authors reviewed the final manuscript.

\section*{Data Availability}

The clinical data used for the training, validation, and test sets were collected at Byers Eye Institute, Stanford School of Medicine (Palo Alto, CA, United States) and was used in a de-identified format.
They are not publicly available and restrictions apply to their use.
The external test datasets were obtained from the Chinese University of Hong Kong (Hong Kong), Narayana Nethralaya (Bangalore, India), and Tilganga Institute of Ophthalmology (Kathmandu, Nepal), and are subject to the respective institutional and national ethical approvals.

\bibliography{refs}
\bibliographystyle{ieeetr}

\clearpage

\section*{Supplementary Material}

\renewcommand{\thesubsection}{S\arabic{subsection}}
\renewcommand{\thetable}{S\arabic{table}}
\renewcommand{\thefigure}{S\arabic{figure}}
\setcounter{figure}{0}
\setcounter{table}{0}

\subsection{Supplementary Tables}

\begin{table}[htb!]
\centering
\caption{Criteria used for labeling cases in the Stanford dataset as \textit{True Glaucoma} and \textit{True Normal}.}
\label{suptbl:criteria}
\begin{tabularx}{\textwidth}{l<{\hsize=0.45\hsize}X<{\hsize=0.45\hsize}X} \toprule
\textbf{Labels} & \multicolumn{1}{c}{\textbf{True Glaucoma}} & \multicolumn{1}{c}{\textbf{True Normal}} \\ \midrule
\multirow{19}*{\textbf{Criteria}} &
\begin{itemize}[leftmargin=*]
    \item Clinical Glaucomatous Disc changes (as per ISGEO classification \cite{foster2002definition}), \textit{and}
    \item OCT Glaucomatous defects on deviation maps and not all green on OCT RNFL and/or OCT GCIPL maps, \textit{and}
    \item 2 repeatable VF defects as per HAP criteria \cite{anderson1992automated}. Reliably measured data were used, \textit{i.e.}~with a fixation loss $<20\%$, false positive errors $<15\%$, and false negative errors $<33\%$, \textit{or} total cupping of the optic nerve and unable to perform VF evaluation, \textit{and}
    \item On Treatment for Glaucoma or has undergone surgery/SLT-ALT.
\end{itemize} &
\begin{itemize}[leftmargin=*]
    \item No disc changes for glaucoma (few cases have high cup disc ratio $> 0.6$ but no other glaucomatous disc changes), \textit{and}
    \item No OCT glaucomatous defects on deviation maps and all green OCT RNFL and OCT GCIPL maps, \textit{and}
    \item No visual field defects, \textit{and}
    \item No treatment/review after a duration no lesser than a year as per chart review.
\end{itemize} \\ \bottomrule
\end{tabularx}
\end{table}

\begin{table}[htb!]
\centering
\caption{Demographic background of the training set from Stanford.
Significance tests for the \textit{True Normal} subset are performed relative to the \textit{True Glaucoma} subset of the Stanford training set.
\textbf{Mean Deviation} (MD) is an overall value of the total amount of visual field loss.}
\label{suptbl:stanfordtrainingdemography}
\begin{tabular}{@{}lcc@{}} \toprule
      & \textbf{True Glaucoma} & \textbf{True Normal} \\ \midrule
Age   & 69.41 ($\pm14.70$)         & 61.84 ($\pm15.20, p < 0.005$)      \\
Asian & 163 (39.9\%)                & 118 (49.0\%)              \\
Caucasian & 147 (36.0\%)            & 77 (32.0\%)               \\
African American & 15 (3.6\%)      & 10 (4.1\%)               \\
Hispanic & 32 (7.8\%)              & 19 (7.9\%)               \\
Data of ethnicity unavailable & 50 (12.2\%)  & 14 (5.8\%)      \\
Average MD  & -9.75 ($\pm7.50$)        & -0.79 ($\pm1.20$)    \\
Mean Refractive Error & -3.57 ($\pm3.37$)  & -2.20 ($\pm4.62, p < 0.001$) \\ \bottomrule
\end{tabular}
\end{table}

\begin{table}[htb!]
\centering
\caption{Demographic background of the validation set from Stanford.
Significance tests of the \textit{True Normal} and \textit{True Glaucoma} subsets are performed relative to the \textit{True Normal} and \textit{True Glaucoma} subsets of the Stanford training set, respectively.}
\label{suptbl:stanfordvaldemography}
\begin{tabular}{@{}lcc@{}} \toprule
      & \textbf{True Glaucoma} & \textbf{True Normal} \\ \midrule
Age   & 70.09 ($\pm10.37, p = 0.74$)     & 67.03 ($\pm11.30, p = 0.0715$)      \\
Asian & 14 (25.0\%)                & 13 (43.3\%)              \\
Caucasian & 16 (29.9\%)            & 6 (20.0\%)               \\
African American & 2 (3.6\%)      & 4 (13.3\%)               \\
Hispanic & 6 (11.0\%)              & 2 (6.6\%)               \\
Data of ethnicity unavailable & 17 (31.0\%)  & 6 (20\%)      \\
Average MD  & -7.89 ($\pm4.17, p = 0.0724$)    & -1.31 ($\pm1.06, p = 0.0241$)    \\
Mean Refractive Error & -2.16 ($\pm4.17, p = 0.1949$)  & -0.53 ($\pm1.99, p <0.005$) \\ \bottomrule
\end{tabular}
\end{table}

\begin{table}[htb!]
\centering
\caption{Demographic background of the test set from Stanford.
Significance tests of the \textit{True Normal} and \textit{True Glaucoma} subsets are performed relative to the \textit{True Normal} and \textit{True Glaucoma} subsets of the Stanford training set, respectively.}
\label{suptbl:stanfordtestdemography}
\begin{tabular}{@{}lcc@{}} \toprule
      & \textbf{True Glaucoma} & \textbf{True Normal} \\ \midrule
Age   & 69.82 ($\pm16.15, p = 0.7886$)  & 63.00 ($\pm16.93, p = 0.4838$)      \\
Asian & 60 (38.2\%)                & 57 (50.0\%)              \\
Caucasian & 65 (41.4\%)            & 42 (36.8\%)               \\
African American & 8 (5.0\%)      & 6 (5.2\%)               \\
Hispanic & 13 (8.2\%)              & 4 (3.5\%)               \\
Data of ethnicity unavailable & 11 (7.0\%)  & 5 (4.3\%)      \\
Average MD  & -9.01 ($\pm7.52, p = 0.2709$)   & -0.79 ($\pm0.98, p = 1.000$)    \\
Mean Refractive Error & -2.64 ($\pm2.86, p = 0.0011$)  & -1.92 ($\pm2.03, p = 0.1552$) \\ \bottomrule
\end{tabular}
\end{table}

\begin{table}[htb!]
\centering
\caption{Demographic background of the Hong Kong test set, such as gender and ethnicity distribution, and mean values (standard deviations) for visual field parameter mean deviation (MD) and Mean Refractive error.
Significance tests of the \textit{True Normal} and \textit{True Glaucoma} subsets are performed relative to the \textit{True Normal} and \textit{True Glaucoma} subsets of the Stanford training set, respectively.}
\label{suptbl:hkdemography}
\begin{tabular}{@{}lcc@{}} \toprule
      & \textbf{True Glaucoma} & \textbf{True Normal} \\ \midrule
Age   & 65.90 ($\pm9.30, p < 0.005$)    & 61.05 ($\pm8.50, p = 0.5139$)     \\
Asian & 277 (100\%)                & 196 (100\%)              \\
Average MD  & -8.50 ($\pm6.81, p = 0.035$)  & -0.90 ($\pm1.30, p = 0.3526$)        \\
Mean Refractive Error & -0.85 ($\pm2.57, p < 0.005$)    & -0.51 ($\pm2.15, p < 0.005$) \\ \bottomrule
\end{tabular}
\end{table}

\begin{table}[htb!]
\centering
\caption{Demographic background of the India test set, such as gender and ethnicity distribution, and mean values (standard deviations) for visual field parameter mean deviation (MD) and Mean Refractive error.
Significance tests of the \textit{True Normal} and \textit{True Glaucoma} subsets are performed relative to the \textit{True Normal} and \textit{True Glaucoma} subsets of the Stanford training set, respectively.}
\label{suptbl:indiademography}
\begin{tabular}{@{}lcc@{}} \toprule
      & \textbf{True Glaucoma} & \textbf{True Normal} \\ \midrule
Age   & 63.84 ($\pm11.72, p < 0.005$)         & 54.76 ($\pm14.95, p < 0.005$)    \\
Asian & 173 (100\%)                & 130 (100\%)              \\
Average MD  & -12.74 ($\pm9.22, p < 0.005$)  & -2.10 ($\pm1.30, p < 0.0001$)        \\
Mean Refractive Error & -0.48 ($\pm2.25, p < 0.005$)        & -0.44 ($\pm2.19, p < 0.005$) \\ \bottomrule
\end{tabular}
\end{table}

\begin{table}[htb!]
\centering
\caption{Demographic background of the Nepal test set, such as gender and ethnicity distribution, and mean values (standard deviations) for visual field parameter mean deviation (MD) and Mean Refractive error.
Significance tests of the \textit{True Normal} and \textit{True Glaucoma} subsets are performed relative to the \textit{True Normal} and \textit{True Glaucoma} subsets of the Stanford training set, respectively.}
\label{suptbl:nepaldemography}
\begin{tabular}{@{}lcc@{}} \toprule
      & \textbf{True Glaucoma} & \textbf{True Normal} \\ \midrule
Age   & 45.34 ($\pm17.08, p < 0.005$)         & 39.17 ($\pm12.28, p < 0.005$)    \\
Asian & 184 (100\%)                & 173 (100\%)              \\
Average MD  & -8.30 ($\pm7.04, p = 0.0369$)  & -2.32 ($\pm1.47, p < 0.005$)        \\
Mean Refractive Error & -1.38 ($\pm2.38, p < 0.005$)        & -1.17 ($\pm1.36, p < 0.005$) \\ \bottomrule
\end{tabular}
\end{table}

\begin{table}[htb!]
\centering
\caption{Distribution of cases in terms of Glaucoma severity.
Classification based on Mean Deviation (Severe: MD $\leq-12$, Moderate: $-12 < \text{MD} \leq -6$, Mild: $-6 < \text{MD}$).}
\label{suptbl:severity}
\begin{tabular}{@{}lccccc@{}} \toprule
                  &  \textbf{Stanford} & \textbf{Hong Kong} & \textbf{India} & \textbf{Nepal}  \\ \midrule
\textbf{Severe Glaucoma}   &  28.40\%        & 24.00\%   & 44.80\% & 21.10\% \\
\textbf{Moderate Glaucoma} &  18.93\%        & 26.10\%   & 17.20\% & 22.76\% \\
\textbf{Mild Glaucoma}     &  52.66\%        & 49.70\%   & 37.90\% & 56.10\% \\ \bottomrule
\end{tabular}
\end{table}

\begin{table}[htb!]
\centering
\caption{Comparison of myopia severity (in terms of spherical equivalent) between the Stanford, Hong Kong, India, and Nepal test sets.
\textbf{TG} stands for \textit{True Glaucoma} and \textbf{TN} stands for \textit{True Normal}.
Chi-squared test was used for severe myopia distribution analysis (Myopia severity distribution: Severe:  $D\leq-6$, Moderate: $-6 < D \leq -3$, Mild: $-3 < D$, where $D$ is diopter).
Emmetropia is defined as spherical equivalent of $-0.25D$ to $+0.25D$.}
\label{suptbl:myopiaseverity}
\begin{adjustbox}{width=\textwidth,center}
\begin{tabular}{@{}>{\raggedright}p{0.2\textwidth}>{\centering}p{0.2\textwidth}>{\centering}p{0.1\textwidth}>{\centering}p{0.1\textwidth}cc@{}} \toprule
\multirow{2}*{\centering \textbf{Subset}} &
\multirow{2}*{\textbf{Severe Myopia}} &
\textbf{Moderate Myopia} &
\textbf{Mild Myopia}     &
\multirow{2}*{\textbf{Emmetropia}}      &
\multirow{2}*{\textbf{Hypermetropia}}   \\ \midrule
\textbf{Stanford (TG)} & 8.88\% ($p = 0.70$) & 8.10\% & 42.20\% & 11.11\% & 20.00\% \\
\textbf{Stanford (TN)} & 4.20\% ($p = 0.98$) & 10.08\% & 31.09\% & 5.88\% & 47.89\% \\
\textbf{Hong Kong (TG)} & 4.70\% ($p = 0.12$) & 12.50\% & 37.50\% & 5.90\% & 39.20\% \\
\textbf{Hong Kong (TN)} & 0.0\% ($p < 0.001$) & 21.01\% & 15.70\% & 10.50\% & 47.30\% \\
\textbf{India (TG)} & 0.0\% ($p < 0.001$) & 3.94\% & 43.20\% & 22.30\% & 38.10\% \\
\textbf{India (TN)} & 0.0\% ($p < 0.001$) & 16.60\% & 30.30\% & 15.15\% & 37.87\% \\
\textbf{Nepal (TG)} & 2.50\% ($p < 0.001$) & 14.28\% & 43.80\% & 10.70\% & 25.00\% \\
\textbf{Nepal (TN)} & 0.0\% ($p < 0.001$) & 6.38\% & 53.00\% & 0.0\% & 40.40\% \\ \bottomrule
\end{tabular}
\end{adjustbox}
\end{table}

\begin{table}[htb!]
\caption{Comparison of additional clinical data between the primary set and four external evaluation datasets.
The statistical analysis was performed with the MedCalc Software (Version 19.4).
Results are expressed as mean ($\pm$ standard deviation) and paired Student's t-test was used to evaluate the level of significance.
A p-value of 0.005 or less was considered significant.
Chi square test was used for comparisons of categorical demographic data for proportions.
\textbf{TG} stands for \textit{True Glaucoma} and \textbf{TN} stands for \textit{True Normal}.
$n$ indicates the number of eyes in each set.
p-values for Stanford Training (TN) have been computed against Stanford Training (TG).
For all other datasets, TN subset have been compared against Stanford Training (TN), and TG subset have been compared against Stanford Training (TG), respectively.
\textbf{Visual Field Index} (VFI) is a global metric that assigns a number between 1-100 percent based on aggregate percentage of visual function with 100\% being perfect age-adjusted visual field.
\textbf{Pattern Standard Deviation} (PSD) depicts focal defects on visual fields by comparing the differences between the adjacent points on the visual field.}
\label{suptbl:additional}
\begin{adjustbox}{width=\textwidth,center}
\begin{tabular}{@{}>{\raggedright}p{0.3\textwidth}>{\centering}p{0.2\textwidth}>{\centering}p{0.15\textwidth}>{\centering}p{0.2\textwidth}>{\centering}p{0.15\textwidth}c@{}} \toprule
\multirow{2}*{\centering \textbf{Subset}} &
\multirow{2}*{\textbf{Cup-Disc Ratio}} &
\multirow{2}*{\textbf{IOP}} &
\textbf{Gender Distribution (F:M)}     &
\multirow{2}*{\textbf{PSD}}      &
\multirow{2}*{\textbf{VFI}}   \\ \midrule

\textbf{Stanford Training (TG)} \\ $(n = 363)$ & \multirow{2}*{$0.80~(\pm 0.12)$}  & \multirow{2}*{$20.07~(\pm 4.75)$} & \multirow{5}*{55:45} & \multirow{2}*{$7.71~(\pm 6.66)$} & \multirow{2}*{74.40\%} \\

 & & & & & \\

\textbf{Stanford Training (TN)} \\ $(n = 291)$ & $0.46~(\pm 0.16)$ $p < 0.005$ & \multirow{2}*{$15.67~(\pm 2.72)$} &  & \multirow{2}*{$1.83~(\pm 0.53)$} & \multirow{2}*{98.46\% $(p < 0.005)$} \\ \midrule

\textbf{Stanford Validation (TG)} \\ $(n = 48)$ & $0.77~(\pm 0.13)$ $p = 0.0856$ & $19.60~(\pm 4.80)$ $p = 0.492$ & \multirow{5}*{49:51 ($p = 0.3240$)} & $7.73~(\pm 4.30)$ $p= 0.982$ &
{77.70\% $(p = 0.5971)$} \\

 & & & & & \\

\textbf{Stanford Validation (TN)} \\ $(n = 39)$ & $0.53~(\pm 0.20)$ $p = 0.0422$ & $14.20~(\pm 3.89)$ $p = 0.01$ &  & $1.72~(\pm 2.46)$ $p = 0.54$ & \multirow{2}*{98.30\% $(p = 0.9468)$} \\ \midrule

\textbf{Stanford Test (TG)} \\ $(n = 157)$ & $0.79~(\pm 0.19)$ $p = 0.5262$  & $19.56~(\pm 5.47)$ $(p = 0.244)$ & \multirow{5}*{49:51 ($p = 0.2194$)} & $6.37~(\pm 4.46)$ $(p=0.0205)$ & \multirow{2}*{77.01\% ($p = 0.6191$)} \\

 & & & & & \\

\textbf{Stanford Test (TN)} \\ $(n = 113)$ & $0.45~(\pm 0.16)$ $p = 0.4764$ & $16.00~(\pm 2.72)$ $(p=0.153)$ &  & $1.12~(\pm 1.07)$ $(p < 0.005)$ & \multirow{2}*{98.06\% $(p = 0.7678)$} \\ \midrule

\textbf{Hong Kong (TG)} \\ $(n = 277)$ & \multirow{2}*{No Data} & $16.19~(\pm 4.17)$ $(p < 0.005)$ & \multirow{5}*{67:33 $(p < 0.005)$} & $6.44~(\pm 4.21)$ $(p < 0.005)$ & \multirow{2}*{79.83\% $(p = 0.5239)$} \\

& & & & & \\

\textbf{Hong Kong (TN)} \\ $(n = 196)$ & \multirow{2}*{No Data} & $13.44~(\pm 2.72)$ $(p < 0.005)$ &  & $1.46~(\pm 0.30)$ $(p < 0.005)$ & \multirow{2}*{99.61\% $(p = 0.2346)$} \\
& & & & & \\ \midrule

\textbf{India (TG)} \\ $(n = 171)$ & \multirow{2}*{No Data} & \multirow{2}*{No Data} & \multirow{5}*{40:60 $(p < 0.005)$} & $7.68~(\pm 3.81)$ $(p = 0.951)$ & \multirow{2}*{65.38\% $(p = 0.0331)$} \\

 & & & & & \\

\textbf{India (TN)} \\ $(n = 121)$ & \multirow{2}*{No Data} & \multirow{2}*{No Data} &  & $2.54~(\pm 1.39)$ $(p < 0.005)$ & \multirow{2}*{93.17\% $(p = 0.006)$} \\
& & & & & \\ \midrule

\textbf{Nepal (TG)} \\ $(n = 166)$  & \multirow{2}*{No Data} & $16.56~(\pm 4.74)$ $(p < 0.005)$ & \multirow{5}*{40:60 $(p < 0.005)$} & $5.37~(\pm 3.30)$ $(p < 0.005)$ & \multirow{2}*{77.00\% $(p = 0.5791)$} \\

 & & & & & \\

\textbf{Nepal (TN)} \\ $(n = 181)$ & \multirow{2}*{No Data} & $15.68~(\pm 2.90)$ $(p = 0.972)$ &  & $1.99~(\pm 1.08)$ $(p = 0.051)$ & \multirow{2}*{97.58\% $(p = 0.5362)$} \\
& & & & & \\ \bottomrule
\end{tabular}
\end{adjustbox}
\end{table}

\clearpage

\subsection{Supplementary Figures}

\begin{figure}[htb!]
\centering
\includegraphics[width=\linewidth]{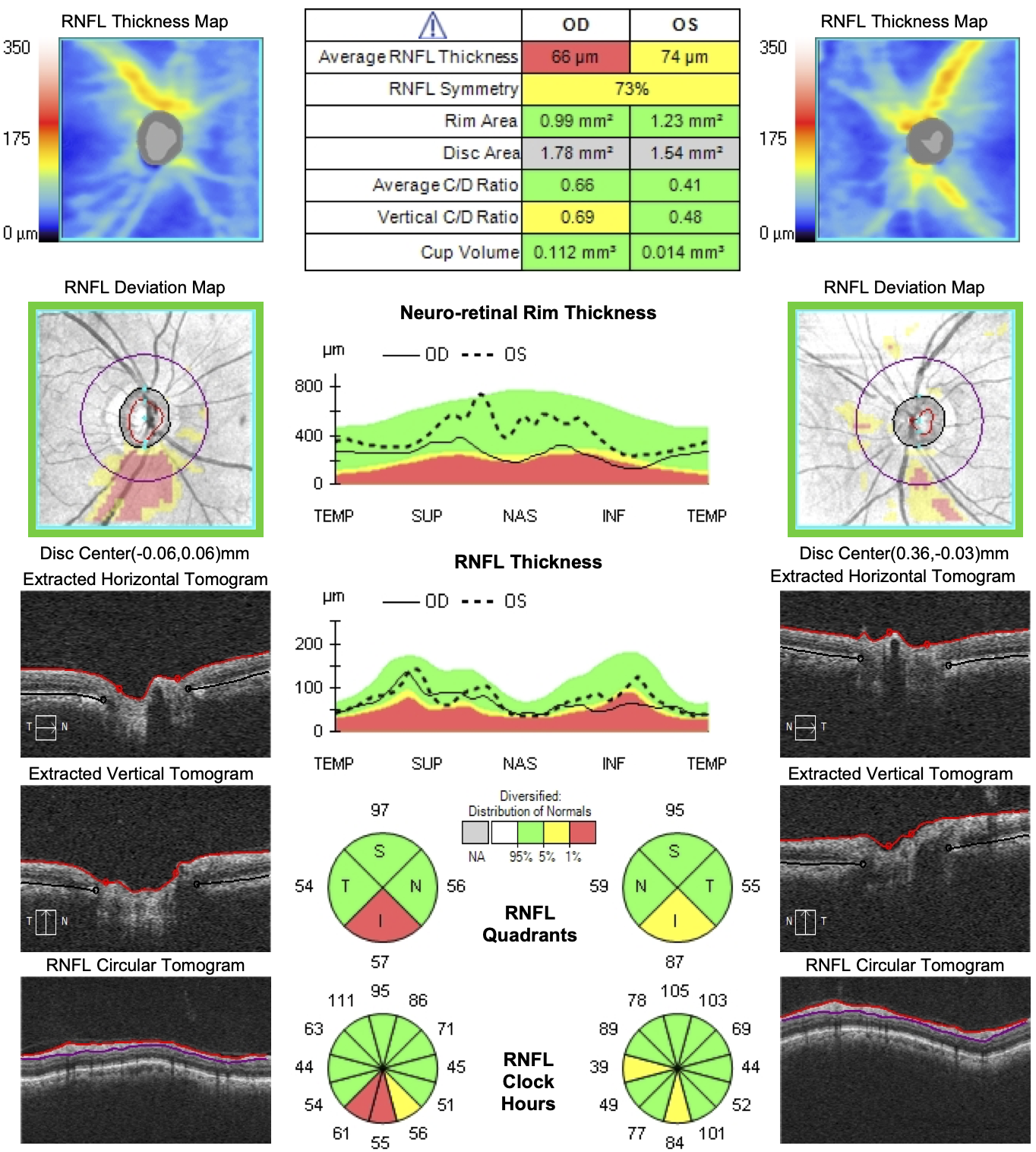}
\caption{Cirrus OCT RNFL and ONH Analysis Map.
\textbf{RNFL thickness} map is a topographical display of RNFL.
\textbf{RNFL Deviation Map} shows deviation from the normal.
OCT en face fundus image depicts boundaries of the cup and disc and the RNFL calculation circle at 3.4mm.
\textbf{Neuro-retinal Rim Thickness} is matched to normative data.
RNFL TSNIT graph displays RNFL measurement along the calculation circle, compared to normative data.
\textbf{RNFL Quadrant} and \textbf{Clock Hour} average thickness is matched to normative data.
Horizontal and vertical B-scans are extracted from the data cube through the center of the disc.
For a particular age and disc size, the patient is expected to have rim volume, C/D ratio, rim area, and cup volume within certain ranges; depicted in a table format which is color coded red, yellow, green, and white, based on how they compare to normal ranges~\cite{meditec2011cirrus}.
Note that some feature based segmentation model has been applied to the raw data to help clinicians be able to interpret the raw cube data.}
\label{supfig:rnflmap}
\end{figure}

\end{document}